\documentclass[pdftex]{aa}  
\usepackage{graphicx}
\usepackage{txfonts}
\usepackage{natbib}
\bibpunct{(}{)}{;}{a}{}{,}	
\begin{document}

   \title{High-dynamic-range extinction mapping of infrared dark clouds}
	   \subtitle{Dependence of density variance with sonic Mach number in molecular clouds}

   \author{J. Kainulainen	\inst{1} \& 
           J.~C. Tan		\inst{2}}

   \offprints{jtkainul@mpia.de}

   \institute{Max-Planck-Institute for Astronomy, K\"onigstuhl 17, 69117 Heidelberg, Germany \\
              \email{jtkainul@mpia.de} \and
               Departments of Astronomy \& Physics, University of Florida, Gainesville, FL 32611, USA \\
              \email{jt@astro.ufl.edu}
            }
   \date{Received ; accepted }
  \abstract
   {Measuring the mass distribution of infrared dark clouds (IRDCs) over the wide dynamic range of their column densities is a fundamental obstacle in determining the initial conditions of high-mass star formation and star cluster formation.}
   {We present a new technique to derive high-dynamic-range, arcsecond-scale resolution column density data for IRDCs and demonstrate the potential of such data in measuring the density variance - sonic Mach number relation in molecular clouds.}
   {We combine near-infrared data from the \emph{UKIDSS/Galactic Plane Survey} with mid-infrared data from the \emph{Spitzer/GLIMPSE} survey to derive dust extinction maps for a sample of ten IRDCs. We then examine the linewidths of the IRDCs using  $^{13}$CO line emission data from the \emph{FCRAO/Galactic Ring Survey} and derive a column density - sonic Mach number relation for them. For comparison, we also examine the relation in a sample of nearby molecular clouds.}
   {The presented column density mapping technique provides a very capable, temperature independent tool for mapping IRDCs over the column density range equivalent to $A_\mathrm{V} \simeq 1-100$ mag at a resolution of $2\arcsec$. Using the data provided by the technique, we present the first direct measurement of the relationship between the column density dispersion, $\sigma_{N / \langle N \rangle}$, and sonic Mach number, ${\cal M}_\mathrm{s}$, in molecular clouds. We detect correlation between the variables with about 3-$\sigma$ confidence. We derive the relation $\sigma_{N / \langle N \rangle} \approx (0.047 \pm 0.016)  {\cal M}_\mathrm{s}$, which is suggestive of the correlation coefficient between the volume density and sonic Mach number, $\sigma_{\rho / \langle \rho \rangle} \approx (0.20^{+0.37}_{-0.22})  {\cal M_\mathrm{s}}$, in which the quoted uncertainties indicate the 3-$\sigma$ range. When coupled with the results of recent numerical works, the existence of the correlation supports the picture of weak correlation between the magnetic field strength and density in molecular clouds (i.e., $B \propto \rho^{0.5}$). While our results remain suggestive because of the small number of clouds in our demonstration sample, the analysis can be improved by extending the study to a larger number of clouds.}
   {}
  \authorrunning{J. Kainulainen et al.}
  \titlerunning{High-dynamic-range column density data for infrared dark clouds}
  \maketitle


\section{Introduction} 
\label{sec:intro}


Measuring the structure and dynamics of the dense, molecular
interstellar medium (ISM) is of crucial importance for our
understanding of star formation from low- to high-mass stars 
and star clusters. 
%
There are a number of outstanding open questions in star formation that come down to the coupling between star formation and the fundamental physical processes acting in molecular clouds. What determines the star-formation rates and efficiencies in molecular clouds and why are those parameters generally low? What is the relative importance of gravity, turbulence, and magnetic fields during molecular cloud evolution and does this vary for low- and high-mass star formation? These questions can be approached by studying the physical characteristics of molecular clouds at different evolutionary stages. Of especial importance are studies of clouds that have not yet started to form stars or are in a very early stage of doing so. This is because such clouds still bear the imprints of the processes that initially formed them and are responsible for driving them toward star formation. For reviews of these topics, we refer to \citet{mck07}.


Our knowledge on the maternal structure of clouds potentially forming high-mass stars ($M \gtrsim 8$ M$_\odot$) and star-clusters is relatively scarce. 
%
%
This is, in part, because of the rapid destruction of the maternal cloud structure by stellar winds, outflows, ionization, and radiation pressure from the first high-mass stars to form in the cloud. Once the presence of high-mass stars in a cloud can be observationally confirmed, the maternal cloud structure has already been heavily affected by them. In part, the difficulties also originate from trivial observational restrictions: most sites that are thought to be favorable for future high-mass star and star cluster formation are located at distances of $D \gtrsim 1.5$ kpc, and they are concentrated in the Galactic plane. This makes attaining sensitive, high-spatial resolution data on their structure a very challenging task. 


Only relatively recently, a class of dense molecular cloud clumps now commonly referred to as infrared dark clouds (IRDCs, hereafter) was discovered \citep[][]{per96, ega98}. These
objects are currently regarded as the best candidates of future high-mass star formation \citep[e.g.,][]{car00, gar04, beu05, rat05, rat06}, although how commonly they might form massive stars remains still unclear \citep[e.g.,][]{kau10}. In the most recent years, the study of IRDCs has seen the coming of large-area Galactic plane surveys that examine the Galactic-scale distribution and properties of IRDCs through comprehensive, statistical samples \citep[e.g.,][]{sim06, jac08, per10, agu11}. 


One persistent obstacle in defining the physical conditions in the maternal environment of IRDCs is our inability to measure, in a uniformly defined manner, the mass distribution over the wide dynamic range of (column-/mass surface-) densities present in IRDCs. Various observational techniques are commonly used in probing the mass distributions: CO emission line observations \citep[e.g.,][]{rom10}, thermal dust emission at (sub)mm wavelengths \citep[e.g.,][]{sch09, mol10}, dust extinction mapping using background stars in near-infrared \citep[NIR, e.g.,][]{kai11a}, and dust extinction mapping in mid-infrared (MIR) using either point source data \citep[e.g.,][]{ryg10} or surface brightness data \citep[e.g.,][]{but09, rag09, vas09}. 
%
%
All these mass-tracing techniques have their specific properties that restrict the parameter space (usually, column density range) within which they can provide accurate mass measurements. For example, CO line emission is known to be linearly correlated with column density only in a very narrow column density\footnote{We note that column density (and mass surface density) are generally related to dust extinction, $A_\mathrm{V}$, through a linear relationship. To make comparisons with relevant earlier works easier, we use in this paper equivalent $A_\mathrm{V}$ as the main unit for column density. We give the detailed conversions to column density and mass surface density in Sections \ref{subsec:nirdata}-\ref{subsec:mirdata}.}
range ($A_\mathrm{V} \approx 3-6$ mag) \citep[e.g.,][]{goo09}. Dust emission measurements depend on the temperature of the dust grains that can vary by a factor of $\sim 2$ in the clouds \citep[e.g.,][]{bon10, stu10, arz11} and of their emissivity index. Dust extinction measurements in the near-infrared are limited to relatively low column densities, typically to $A_\mathrm{V} \approx 1-25$ mag in \citep[e.g.,][]{lom01, kai11a}. For MIR surface brightness extinction mapping, higher column densities can be probed up to $A_\mathrm{V} \approx 100$ mag, but, as discussed below, systematic uncertainties limit the accuracy below $A_\mathrm{V} \approx 10$ mag. As a result, it is not straightforward to relate the mass distribution at small, dense, star-forming scales to the surrounding larger-scale, more diffuse environment.


In our previous works, we have developed techniques to probe the mass distributions of IRDCs by measuring the NIR \citep{kai11a} and MIR \citep[][BT09 and BT12, hereafter]{but09, but12} dust extinction through them. While the NIR technique has proven to be relatively sensitive, reaching column densities of $A_\mathrm{V} \approx 25$ mag at a resolution of $\sim 30\arcsec$, the MIR data probe reliably column densities between $A_\mathrm{V} \simeq 10-100$ mag at a resolution of $\sim 2\arcsec$. Both of the techniques are essentially based on a model for dust optical properties in infrared which are relatively well-known, and the dynamical ranges probed by them clearly overlap. This provides a good basis for a conception of combining the two techniques with aim of achieving both extended dynamical range and high spatial resolution. This goal is further motivated by the fact that the MIR technique is known to have caveats affecting especially low-column densities. The technique effectively filters out the large-scale, diffuse ($A_\mathrm{V} \lesssim 5-10$ mag) extinction component, because of the estimation of the background intensity \citep[for details, see][BT09 and BT12 hereafter, respectively]{but09, but12}. The details of the adopted filtering scheme can significantly affect the performance of the technique \citep[e.g., BT09][]{wil12}. However, the diffuse dust component that is filtered away is well recovered by the NIR data, and it would seem feasible to use the NIR data to "re-calibrate" the MIR data, thus  alleviating the uncertainty of the MIR technique arising from the large-scale filtering needed to estimate the MIR background intensity.


We develop in this paper a new technique for probing the mass distributions of IRDCs that takes the advantage of both NIR point source data and MIR surface brightness data. The technique is based on data that are readily available for a large portion of the Galactic plane, i.e., NIR data from the \emph{UKIDSS/Galactic Plane Survey} \citep{luc08} and MIR data from the \emph{Spitzer/GLIMPSE} survey \citep{ben03}. 
We will demonstrate the use of the technique by deriving column density data for ten IRDCs and their surroundings. 


We also demonstrate in this paper the scientific potential of high-dynamic-range column density data in examining the probability density functions (PDFs) of column densities in the clouds. The column density PDFs can be used to study the roles of gravity and turbulence in molecular clouds \citep[e.g.,][]{fro07, goo09, kai09, fro10}, generally building on the fact that simulations of supersonic turbulent media predict a log-normal shape for the \emph{density} PDF \citep[e.g.,][]{vaz94, pad97mnras, bal11, kri11}
\begin{equation}
p(x) = \frac{1}{ \sigma_\mathrm{\ln{x}} \sqrt{2\pi} x } \exp{\big[-\frac{(\ln{x}- \mu )^2}{2 \sigma_\mathrm{\ln{x}}^2}\big]},
\label{eq:lognormal}
\end{equation}
where $x=\rho/\rho_0$ is the mean-normalized volume density and $\mu$ and $\sigma_\mathrm{\ln{x}}$ are the mean and standard deviation in logarithmic units. Deviations from the log-normal shape are predicted for strongly gravitating systems, inducing higher relative fraction of high-column density material \citep[e.g.,][]{kle00, fed08a}. In K09 we presented an analysis of the PDFs of a sample of nearby molecular clouds and concluded that the PDFs of star-forming clouds are consistent with a log-normal (or normal) shaped function at lower column densities ($A_\mathrm{V} \lesssim 2-5$ mag) and a power-law like extension at higher columns ($A_\mathrm{V} \gtrsim 2-5$ mag). In contrast, the PDFs of non-star-forming clouds were  usually very well described with log-normals. This led us to suggest that quiescent clouds are primarily shaped by turbulence, while the effect of self-gravity is evident in the PDFs of star-forming clouds. However, in \citet{kai11b} we analyzed the data further and suggested that the PDF tail could be related to clumps that are not self-gravitating but rather supported significantly by external pressure from the surrounding, gravitating cloud material.

In this paper, we also use the derived column density data together with CO line emission data to examine the relationship between density fluctuations, measured by the column density dispersion $\sigma_{N / \langle N \rangle}$, and turbulence energy, measured by the sonic Mach number ${\cal M\mathrm{s}}$, in the clouds. These variables are expected to be correlated in turbulent media and a correlation coefficient between them is assumed by numerous analytic models of star formation \citep[e.g.,][]{pad97mnras, pad02, kru05, hen08, elm11}. We measure the relationship between ${\cal M_\mathrm{s}}$ and $\sigma_{N / \langle N \rangle}$ in the IRDCs and also in a larger sample that is supplemented by including a set of nearby molecular clouds we have previously derived column density data for \citep[][hereafter K09]{kai09}. Our results demonstrate that high-dynamic-range data can, and are in fact required to, probe the relation.

\section{Data}           
\label{sec:data}

\subsection{The IRDC sample}
	
We use as a sample the ten IRDCs selected by BT09. The coordinates of the clouds are listed in Table \ref{tab:clouds}. The clouds were originally chosen from the sample of \citet{rat06} as massive, relatively nearby clouds with high contrast against the Galactic background, being thereby well-suited for the 8 $\mu$m extinction mapping. The clouds are distributed between the Galactic longitudes of $l = [18^\circ, 53^\circ]$ and kinematic distances of $D \approx 1.8 - 5.7$ kpc \citep[derived by][]{sim06}. The most massive cloud (D) has a MIR-derived mass of $M_\mathrm{MIR} = 5.34 \times 10^4$ M$_\odot$. The masses are listed in Table \ref{tab:clouds}. The masses were derived by BT12 and they comprise the material inside the ellipsoids defined by \citet{sim06}, which is likely to be $\sim10$\% of the total mass of the clouds estimated over larger scales using NIR extinction mapping \citep{kai11a}. Thus, the most massive clouds in the sample are somewhat more massive than the nearby Orion A molecular cloud ($M \approx 10^5$ M$_\odot$, e.g., K09). All clouds contain mm-cores at $11\arcsec$ resolution \citep{rat06}, most of which also appear as single cores at the $2\arcsec$ resolution of the MIR dust extinction data (BT12). The masses of the cores are typically some tens of solar masses, but depending on how much of the surrounding envelope is included, the masses can also be higher by a factor of a few (BT12). While it is currently not well-constrained how massive a core is needed for high-mass star formation, the cores in principle contain enough material to form such stars.

\begin{table*}
\begin{minipage}[t]{\textwidth}
\caption{Properties of the IRDCs.}             
\label{tab:clouds}      
\centering          
\renewcommand{\footnoterule}{}  
\begin{tabular}{l c c c c c c c c c c c c}
\hline\hline
\#\footnote{Naming according to BT09.} & $l$  & $b$  & $D\footnote{From \citet{sim06}.}$ & $\sigma_\mathrm{v}$ & $\sigma_\mathrm{v}$ & $\sigma_\mathrm{v}$ & $\sigma_\mathrm{v}$ & $\sigma_{N / \langle N \rangle}$ & $\Sigma _\mathrm{sat}$\footnote{Saturation limit, i.e., the maximum observable mass surface density, of the mid-infrared data (cf., BT12).}  & $M_\mathrm{MIR}$ & $M_\mathrm{combined}$  & $\alpha_\mathrm{vir}$	\\
 	&		&		&	 & (G, box)\footnote{The letters \emph{G}(aussian) and \emph{F}(ixed range) refer to the two approaches used in estimating the velocity dispersions in the clouds (see Section \ref{subsec:co}). The abbreviations \emph{box} and \emph{ell} refer to the two approaches used in defining the cloud area (see Section \ref{subsec:area}).  }	& (F, box)$^d$	& (G, ell)$^d$		& (F, ell)$^d$	& (box)$^d$	&	&	&   &  \\
 	& [$^\circ$] & [$^\circ$] & [kpc] & [km s$^{-1}$] & [km s$^{-1}$] & [km s$^{-1}$] & [km s$^{-1}$] &  & [g cm$^{-2}$] & [$10^3$ M$_\odot$] & [$10^3$ M$_\odot$] & \\
\hline
A 	& 18.822	& -0.285 	& 4.8 &  2.04		& 3.44	& 2.05		& 3.78	& 0.63		& 0.496 &  17.7 & 18.5  & 	1.4 \\
B 	& 19.271	& 0.074	& 2.4 & 1.60		& 1.71	& 1.77		& 1.79	& 0.69		& 0.488 &  2.2  & 2.2	& 2.2 \\
C  	& 28.373	& 0.076	& 5.0 & 3.72		& 3.38	& 3.75		& 3.41	& 0.80		& 0.520 &  45 	& 53.2 & 2.4	\\
D 	& 28.531	& -0.251	& 5.7 & 1.85		& 2.12	& 2.01		& 2.14	& 0.67		& 0.436 &  53.4 & 74.3\footnote{$M_\mathrm{combined}$ can be somewhat underestimated because of incomplete coverage of the NIR data, see Fig. \ref{fig:maps-D}.} & 0.5	 	\\
E 	& 28.677	& 0.132	& 5.1 & 4.32 		& 1.94	& 2.15		& 2.55	& 0.60 		& 0.504 & 25.2  & 28.7 &1.1 \\
F 	& 34.437	& 0.245	& 1.56\footnote{The distance for the cloud F has been recently derived from parallax measurements \citep{kur11} and we adopt this distance  \citep[however, see also][]{fos12}. For comparison, the $GRS$ distance for the cloud is $D = 3.7$ kpc).} & 3.62		& 4.15	& 1.86		& 2.54	& 0.73		& 0.370 &  0.79\footnote{The mass given by BT12 was scaled to correspond the adopted distance.}  & 0.85 & 3.5	\\
G 	& 34.771	& -0.557	& 2.9 & 3.28		& 3.15	& 2.95		& 2.90	& 0.57		& 0.347 & 2.01   & 3.3 & 4.7\\
H 	& 35.395	& -0.336	& 2.9 & 2.03		& 2.11	& 1.48		& 1.78	& 0.52		& 0.416 & 13.34 & 16.7 & 0.7	   	\\
I 	& 38.952	& -0.475	& 2.7 & 1.65		& 1.72	& 1.26		& 1.32	& 0.63		& 0.402 & 2.05 	 & 2.7  &   1.2 \\
J 	& 53.116	& 0.054	& 1.8 & 0.96		& 1.14	& 0.83		& 0.95	& 0.50		& 0.328 &  0.08 & 0.2   &  1.5 \\
\hline                  
\end{tabular}
\end{minipage}
\end{table*}

\subsection{NIR dust extinction data}
\label{subsec:nirdata}


We use in this work NIR dust extinction data as a tracer of low-extinction material ($A_\mathrm{V} < 10$ mag) in the IRDCs. The data were derived using the NIR color-excess mapping scheme presented by \citet{kai11a}, which is based on the so-called \emph{NICER} technique \citep{lom01}, but refined to deal with the diffuse extinction component in the Galactic plane and with the large number of foreground sources towards IRDCs. For the detailed description of the technique, we refer to \citet{kai11a} and \citet{lom01}. \citet[][see their Appendix A]{kai11a} presented the NIR extinction maps for the clouds A, B, G, H, and I. The maps for the remaining clouds were computed for this work. 

The NIR extinction mapping technique was used in conjunction with the $JHK_\mathrm{S}$ band data from the \emph{UKIDSS/Galactic Plane Survey/Data Release 6 Plus}. The UKIDSS project is defined in \citet{law07}. \emph{UKIDSS} uses the \emph{UKIRT Wide Field Camera} \citep[\emph{WFCAM;}][]{cas07} The photometric system is described in \citet{hew06}, and the calibration is described in \citet{hod09}. The science archive is described in \citet{ham08}. The derived NIR extinction maps have the sensitivity (statistical uncertainty) of $\sigma (A_\mathrm{V}) \approx 0.4-0.7$ mag, depending somewhat on the Galactic coordinates and the distance of the clouds. The zero-point of the data is rather uncertain because of the uncertainties in the foreground source removal and in estimating the extinction because of the diffuse dust in the Galactic plane; \citet{kai11a} give the value of $\sigma_0 (A_\mathrm{V}) \approx 1$ mag for the zero-point uncertainty. The spatial resolution of the data is $FWHM = 30"$ and at that resolution they typically span extinction values between $A_\mathrm{V} \approx 1-30$ mag.
We note that in derivation of the NIR extinction maps a wavelength dependency of optical depth, i.e., the reddening law, is assumed. \citet{kai11a} used the reddening law as derived by \citet{car89}
\begin{equation}
\tau_\mathrm{K} = 0.600 \tau_\mathrm{H} = 0.404 \tau_\mathrm{J}
\label{eq:cardelli}
\end{equation}


It should be emphasized that the NIR technique determines the extinction in NIR, e.g., $A_\mathrm{K}$. In order to estimate column densities, we convert the NIR extinction to visual extinction using the coefficient from \citet{car89}
\begin{equation}
\tau_\mathrm{K} = 0.114 \tau_\mathrm{V}.
\label{eq:cardelli-v}
\end{equation}
We note that we combine our results later in this paper with K09 who used the same relative extinction law. Then, the calculated visual extinction can be converted to gas column density \citep{sav77, boh78, vuo03}
\begin{equation}
N_\mathrm{H}^\mathrm{NIR} = 2N(\mathrm{H}_2)+N(\mathrm{H}) = 1.9\times 10^{21} \mathrm{\ cm}^{-2}  \big( \frac{A_\mathrm{V}}{\mathrm{mag}} \big).
\label{eq:bohlin}
\end{equation}
Column density can further be expressed in terms of the total mass surface density by adding 40\% (implying the H:He number ratio of 10:1) of helium
\begin{equation}
\Sigma_\mathrm{NIR} = 1.4  m_\mathrm{H}  N_\mathrm{H}^\mathrm{NIR}
\label{eq:Sigma_NIR}
\end{equation}


We can also express the \emph{empirical} conversion from extinction to mass surface density (i.e., Eqs. \ref{eq:cardelli-v}-\ref{eq:Sigma_NIR}) with the help of an "effective" opacity, $\kappa^\mathrm{e}_\mathrm{K}$, through the definition
\begin{equation}
\tau_\mathrm{K} = \kappa^\mathrm{e}_\mathrm{K} \Sigma_\mathrm{NIR}.
\label{eq:kappa}
\end{equation}
Inputing Eqs. \ref{eq:cardelli-v}-\ref{eq:Sigma_NIR} to this results in the effective opacity
\begin{equation}
 \kappa^\mathrm{e}_\mathrm{K} = 25.9 \mathrm{\ cm}^2 \mathrm{\ g}^{-1}.
\end{equation}
Corresponding effective opacities for $H$ and $J$ bands result directly from Eq. \ref{eq:cardelli}
\begin{equation}
 \kappa^\mathrm{e}_\mathrm{H} = 43.2 \mathrm{\ cm}^2 \mathrm{\ g}^{-1},
\end{equation}
\begin{equation}
 \kappa^\mathrm{e}_\mathrm{J} = 64.1 \mathrm{\ cm}^2 \mathrm{\ g}^{-1}.
\end{equation}

\subsection{MIR dust extinction data}
\label{subsec:mirdata}

\begin{figure}
   \centering
\includegraphics[width=0.9\columnwidth]{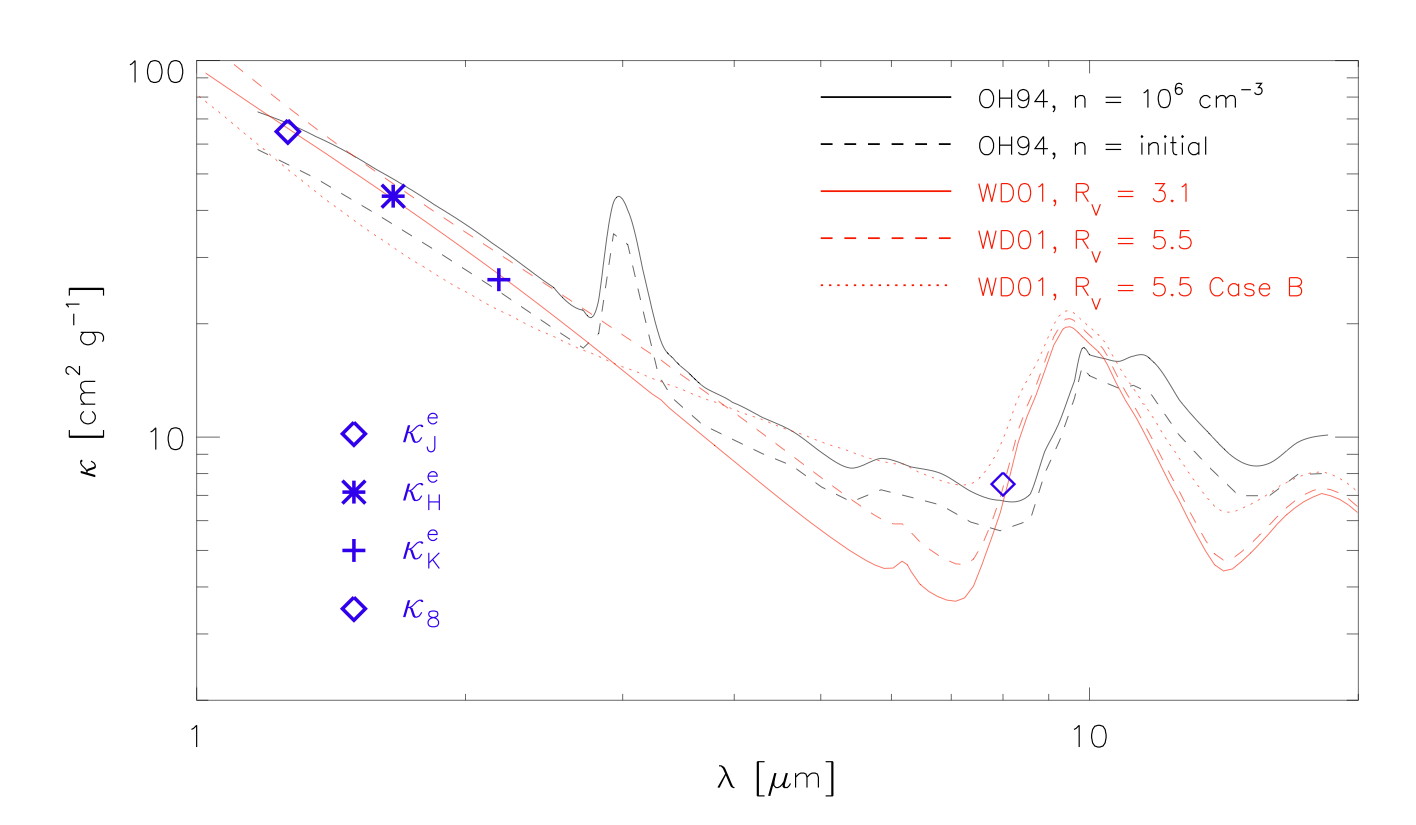} 
      \caption{Opacity laws from \citet{oss94} (black dashed line --- uncoagulated thin ice mantle dust model; black solid line --- thin ice mantle model after coagulation for $10^5$ y at density of $n(\mathrm{H}) = 10^6$ cm$^{-3}$) and \citet{wei01} (red lines, models for $R_\mathrm{V} = 3.1$, $R_\mathrm{V} = 5.5$ and $R_\mathrm{V} = 5.5$ "case B" shown with solid, dashed, and dotted lines). The figure also shows the adopted 8 $\mu$m opacity (diamond) and the effective opacities corresponding to the empirical conversion of NIR extinction to mass surface density (symbols detailed in the panel), see Section \ref{subsec:nirdata}).
              }
         \label{fig:opacitylaws}
   \end{figure}

We use the MIR dust extinction data derived using the \emph{Spitzer/GLIMPSE} survey 8 $\mu$m images by BT12 (see also BT09) as another column-density tracer for the IRDC complexes. The MIR mapping technique is based on measuring the decrease of surface brightness towards the IRDCs, which can be interpreted in terms of 8 $\mu$m optical depth. For the details of the mapping technique we refer to BT12. 
%
%
The derived optical depths are used to estimate mass surface densities, $\Sigma_\mathrm{8}$, with an equation analogous to Eq. \ref{eq:kappa} and adopting a dust opacity at 8 $\mu$m
\begin{equation}
\Sigma_\mathrm{8} = \tau_8 / \kappa_8 .
\end{equation}
BT09 and BT12 chose the value $\kappa_8 = 7.5$ cm$^{2}$ g$^{-1}$, which was taken as an approximate mean of five prospective dust models from \citet{oss94} and \citet{wei01}, but closest to the \citet{oss94} moderately coagulated thin ice mantle dust model (see BT09 for details). BT09 also used weighting by the MIR background spectrum and the filter response function in calculating the opacity values. Figure \ref{fig:opacitylaws} illustrates the opacities $\kappa_8$ and $\kappa^\mathrm{e}_\mathrm{K}$ together with the opacity laws from \citet{wei01} and \citet{oss94}. The figure shows that both in NIR and MIR the adopted opacities represent an average of the shown opacity laws. It is not possible to estimate the accuracy of the adopted 8 $\mu$m opacity, but the standard deviation of the opacities  of the five models is 22\%. Also, it is not straightforward to estimate the uncertainty due to variations in the MIR background spectrum, but if we re-calculate the mean opacity using the case of flat background spectrum, a value $\kappa_8 = 9.0$ cm$^{2}$ g$^{-1}$ follows, indicating 20\% larger value than the mean with the chosen background spectrum. While the flat background spectrum is a rather extreme case, it seems reasonable to use the total uncertainty of $\sim$30\% for the \emph{absolute} value of the MIR opacity. However, one should note that unless average dust grain properties significantly alter from cloud-to-cloud, the relative opacity (or extinction/mass surface density) measurements between clouds is significantly more robust than this value.

The \emph{Spitzer} 8 $\mu$m images have the resolution of $FWHM = 2\arcsec$ ($1\farcs2$ pixel size) and that is also the resolution of the MIR extinction data. At this resolution, the technique probes the mass surface density values up to $\Sigma \lesssim 0.4-0.5$ g cm$^{-2}$, or equivalently, $A_\mathrm{V} \lesssim 92-115$ mag depending on the cloud. These upper limits, or "saturation" values, are estimated empirically as described in BT12 and listed in Table \ref{tab:clouds}. The sensitivity (the absolute uncertainty) of the technique is not straightforward to determine, because the main sources of uncertainty are the variations of the background and foreground surface brightness. BT09 discusses that the technique becomes unreliable at $\Sigma \lesssim 0.013$ g cm$^{-2}$, or $A_\mathrm{V} \lesssim 3$ mag. However, at column densities lower than $\Sigma \lesssim 0.04$ g cm$^{-2}$, or $A_\mathrm{V} \lesssim 9$ mag, the MIR extinction maps show systematic variations that are not present in the NIR data, likely resulting from variations in the MIR background, and hence, from the zero-point uncertainty of the MIR data. The MIR mapping technique is also affected by bright point sources and extended emission (nebulae) in the 8 $\mu$m data. These were masked out during the mapping procedure as detailed in BT09, i.e., we do not estimate extinction in the regions that suffer from significant nebulosity. We also note that the \emph{GLIMPSE} data contains some stripe-like artifacts resulting from non-uniform calibration.


The properties above reflect the complimentary of the NIR and MIR dust extinction techniques. While the NIR data are more sensitive and less affected by systematic uncertainties at $A_\mathrm{V} \lesssim 10$ mag, the MIR data can probe significantly higher column densities. There is also an overlap range between the dynamic ranges of the techniques at about $A_\mathrm{V} \approx 5-15$ mag. 

\subsection{CO line emission data}
\label{subsec:co}


We use the publicly available $^{13}$CO ($J$=1-0) line emission data from the \emph{FCRAO/Galactic Ring Survey} \citep[][hereafter \emph{GRS}]{jac06} to examine the velocity structure of the IRDC complexes. The \emph{GRS} data have a spatial resolution of $FWHM=46\arcsec$, spectral resolution of 0.2 km s$^{-1}$, and typical sensitivity of $\sigma (T_\mathrm{A, rms}) = 0.4 $ K. 


Since the IRDCs are located in the Galactic plane, the lines of sight toward them can show several emission components. We first determined which component corresponds to the IRDC by comparing their morphologies to the extinction data. This was done subjectively by eye, but we note that in almost all cases the choice was trivial; The IRDCs were originally chosen to be in somewhat isolated positions. 
In nine out of ten IRDCs the strongest $^{13}$CO emission peak matched well the morphology of the IRDC as seen by dust extinction. Toward the Cloud E the velocity structure of $^{13}$CO was more complicated (see Fig. \ref{fig:co_box}). Because the dust extinction techniques can probe the total column density up to $D \lesssim 8$ kpc, it appears likely that the extinction data for cloud E is affected by several, physically unrelated clouds. Therefore, we excluded the cloud E from the analyses performed in Section \ref{sec:results}.


We used the $^{13}$CO line emission to estimate the velocity dispersions in the clouds. The estimation was performed inside two regions for each cloud: an ellipsoid as defined by \citet{sim06}, and a box-like region that (often, but not always) contains a closed $A_\mathrm{V} = 7$ mag contour. We describe these regions more closely in Section \ref{subsec:area} and show the average spectra in Appendix \ref{app:co}. For both regions, we estimated the one-dimensional velocity dispersion, $\sigma_\mathrm{v}$, using the following two approaches. The first approach was based on fitting a single Gaussian to the mean spectrum over the velocity interval of the component. The velocity interval was defined using an average spectrum over the region. We calculated the first derivative of the average spectra, and determined the first zero-points of the derivative from both left and right side of the velocity of the maximum intensity. These zero points defined the interval limits. We also required the interval limit to be at most 15 km s$^{-1}$ offset from velocity of the peak. The intervals are indicated in Figs. \ref{fig:co_box} and \ref{fig:co_ell} with vertical dotted lines. Then we fitted a Gaussian to the spectrum within this interval. To get a handle on how much the velocity dispersion depends on the chosen velocity interval, we also fitted a Gaussian to the interval $v = [v_\mathrm{peak} - 1.5 \sigma_\mathrm{v}, v_\mathrm{peak} + 1.5 \sigma_\mathrm{v}]$, where $\sigma_\mathrm{v}$ is the dispersion from the first Gaussian fit. These two velocity dispersions (and the corresponding fits) are also shown in Figs. \ref{fig:co_box} and \ref{fig:co_ell} and we give in Table \ref{tab:clouds} their mean value. The mean value is also used in all further analyses. 

The second approach for calculating the velocity dispersions was calculating the standard deviation of the spectrum in the fixed velocity interval whose definition is described above. As the mean in this calculation, we used the intensity-weighted mean, which represents the center-of-mass in the case where intensity is linearly proportional to the column density along the line of sight. The linear proportionality assumes that the $^{13}$CO emission is optically thin and does not suffer from depletion \citep[see][for a study of these effects in the clouds F and H]{her11tan, her11}. Only channels with intensity higher than three times the root-mean-square noise level were included in the calculation. The resulting velocity dispersions are given in Table \ref{tab:clouds}.


In addition to the \emph{GRS} data, we use in Section \ref{subsec:disc_var} $^{12}$CO data from the \emph{COMPLETE} survey \citep{rid06} and the survey of \citet{dam01} to examine the velocity dispersions of nearby molecular clouds. The \emph{COMPLETE} data was available for Ophiuchus and Perseus. The \citet{dam01} data was adopted for the clouds in the sample of K09 (except for the Lupus clouds and LDN134 that are not covered by Dame et al.). The CO data for these nearby clouds do not suffer from the line-of-sight confusion (they are outside the Galactic plane) and the linewidths toward them were computed from a Gaussian fit to the mean spectrum. In addition, we adopt the mean linewidth for Taurus from \citet{gol08}, i.e., $\sigma (^{13}$CO) $= 1.0$ km s$^{-1}$. 

\section{Methods}           
\label{sec:methods}

\subsection{Combination of NIR and MIR data}
\label{subsec:combination}

\begin{figure}
   \centering
\includegraphics[width=0.9\columnwidth]{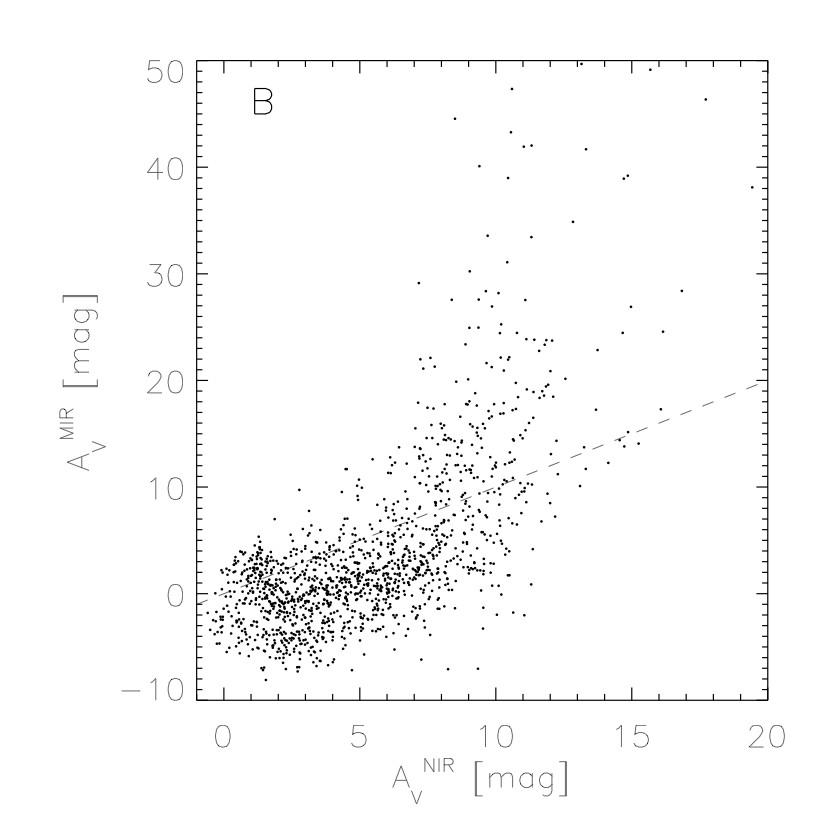} 
      \caption{Pixel-to-pixel comparison of the NIR and MIR column density data for the cloud B. The dashed line shows the one-to-one correlation. The data show correlation below $A_\mathrm{V}^\mathrm{NIR} \lesssim 10-15$ mag, above which the NIR data starts to underestimate extinction compared to the MIR data. The offset of $\sim 5$ mag in the range $A_\mathrm{V}^\mathrm{NIR} \lesssim 10$ mag originates from the spatial filtering of the MIR technique. 
              }
         \label{fig:nir-to-mir-B}
   \end{figure}

We combine the NIR and MIR dust extinction maps with the aim of achieving high-dynamic-range extinction data. 
We first examined the correlation between the data sets by convolving the MIR data ($FWHM=2 \arcsec$) to the significantly lower resolution of the NIR data ($FWHM=30 \arcsec$) and making pixel-to-pixel comparisons. As an example, Fig. \ref{fig:nir-to-mir-B} shows the comparison for the cloud B. Despite the rather high scatter, the data sets correlate at column densities below $A_\mathrm{V} \lesssim 10-15$ mag. In this range, the column densities measured by the MIR data are systematically about $\sim$4 mag lower than the column densities measured from NIR data. The offset is present in all clouds and varies between $A_\mathrm{V} \approx 2-8$ mag. The offset is most likely a result of the spatial filtering of the MIR mapping technique which effectively subtracts the diffuse cloud component from the MIR maps. At column densities higher than $A_\mathrm{V} \gtrsim 10-15$ mag the NIR data begins to underestimate the column density, which results from the fact that the NIR data does not efficiently sample high ($A_\mathrm{V} \gtrsim 30$ mag) extinction lines-of-sight  (none, or very few, background stars are visible through higher-extinction clumps/cores). 


Since the NIR and MIR data showed correlation in the column density range $A_\mathrm{V} \lesssim 10-15$ mag, we conclude that it is reasonable to assume that the difference between the MIR and NIR extinction data below this range is a good representation of the background component that the MIR technique filters out. With this in mind, we combined the NIR and MIR data with the following procedure. First, we re-binned and smoothed the MIR data to the resolution of NIR data ($FWHM=30\arcsec$, pixel size $15\arcsec$) and subtracted it from the NIR image to generate "a correction image"
\begin{equation}
A_\mathrm{V}^\mathrm{correction} \ (30\arcsec) = A_\mathrm{V}^\mathrm{NIR} - A_\mathrm{V}^\mathrm{MIR} (30\arcsec).
\end{equation}
When built in this way, the correction image, $A_\mathrm{V}^\mathrm{correction} \ (30\arcsec)$, approximates the large-scale component filtered out by the MIR technique at spatial scales equal or larger than the background filter size used in the MIR technique (in the case of BT12, a few arcmin). Thus, the correction image clearly resolves the filtering scale of the MIR technique. Since the NIR data underestimates the column density above $A_\mathrm{V} \gtrsim 10-15$ mag, we blanked the pixels in the correction image in which the extinction is higher than $A_\mathrm{V} \geq 10$ mag. Values for these pixels were estimated as a mean within an aperture of $75 \arcsec$ (5 pixels of the NIR data). A Gaussian weighting function with $FWHM=30\arcsec$ was used to weight the values when calculating the mean. Then, the correction image $A_\mathrm{V}^\mathrm{correction} \ (30\arcsec)$ was re-binned to the pixel scale of the MIR data ($1 \farcs 2$) and added to the original MIR column density map
\begin{equation}
A_\mathrm{V}^\mathrm{combined}  = A_\mathrm{V}^\mathrm{NIR} + A_\mathrm{V}^\mathrm{correction} \ (2\arcsec).
\end{equation}
This resulted in column density maps, $A_\mathrm{V}^\mathrm{combined}$, that take advantage of the signals both in the NIR and MIR data. The pixels that lack either NIR or MIR data (e.g., because of the incomplete coverage of the NIR data, bright MIR point sources or extended MIR nebulosity) were flagged to have no data in the maps. As a result, extinction is not determined for every pixel but only for regions in which we consider \emph{both} the NIR and MIR techniques to be unaffected by any artifacts. We describe the results further in Section \ref{subsec:hdr_maps}. In the remainder of this paper, when we refer to dust extinction/column density/mass surface density maps, or data, we refer to these combined maps.

We finally note that the \emph{combination procedure} is not sensitive to the absolute values of the adopted dust opacities (see Fig. \ref{fig:opacitylaws}), only to the \emph{ratio} of dust opacities in NIR and MIR. Only the scaling of the observed extinctions to the units of mass surface density requires knowledge of the absolute dust opacity. In our approach the NIR-MIR dust opacity ratio equals to
\begin{equation}
\frac{\kappa_8}{\kappa^\mathrm{e}_\mathrm{K}} = 0.29 \pm 0.10.
\end{equation}
As shown in Fig. \ref{fig:opacitylaws}, this ratio represents an average behavior of the displayed models. We quote as uncertainty of the ratio the standard deviation of the models used in Fig. \ref{fig:opacitylaws}, which is 30\% for the 8 $\mu$m (see Section \ref{subsec:mirdata}) band and 15\% for the $K$ band. Combining these values quadratically results in uncertainty of 36\%. However, we again point out that this measure of uncertainty reflects a selection of dust models. If the dust grain properties are relatively similar from cloud-to-cloud, the ratio in considerably less uncertain. Currently, there is no detailed information available on this ratio in the particular environment of IRDCs, and hence, aligning the relative reddening law more accurately based on actual measurements is not possible.

\subsection{Definition of the cloud area}
\label{subsec:area}

It is not straightforward to define an object such as "a cloud" in the relatively crowded fields toward the Galactic plane in which the IRDCs reside. In Sections \ref{subsec:moments}-\ref{subsec:disc_var} we examine statistics derived from the column density PDFs of the IRDCs. Despite the seemingly straightforward definition of the PDF, it is not trivial to measure its shape from observational data in a robust way. An ideal target would be a totally isolated cloud in which the lowest column density included in the PDF forms a closed contour in the column density map. Such a closed contour can be regarded as the \emph{completeness limit} of the PDF; The shape of the PDF below the level of the lowest closed contour can clearly suffer from incompleteness. For example, in K09 we used as a definition of a cloud the level $A_\mathrm{V} = 1$ mag, which in most cases was below the mode of the PDF. In the case of the IRDCs that are located in the Galactic plane the confusion from physically unrelated clouds that are nearby in projection makes it more difficult to define the region which should be included in the PDF. 


With this caveat in mind, we adopted two definitions of the cloud area for our analysis. The first definition is to use the ellipsoids defined by \citet{sim06}. We chose from the catalog the ellipsoids with largest area within the mapped area, which in practice roughly encircles the high-column density ($A_\mathrm{V} \gtrsim 10-15$ mag) regions in the IRDCs. The ellipses are shown in the figures of Appendix \ref{app:data}. The second definition of the cloud area was constructed with the aim in being as complete as possible down to relatively low column densities. In the second definition, we defined a rectangular area that surrounds the \citet{sim06} ellipsoid, and makes an effort to completely surround a closed $A_\mathrm{V} = 7$ mag contour. However because of blending of neighboring clouds in two-dimensional data, this contour is not necessarily closed in all clouds. In cases where the blending was severe, we used the velocity structure as traced by the $^{13}$CO data to help disentangling physically unrelated components. The final rectangular regions chosen are shown in the maps in Appendix \ref{app:data}. We conclude that the data within these areas should be close to complete above $A_\mathrm{V} > 7$ mag. However, we also note that in some, more isolated clouds slightly off the Galactic plane, such as clouds C and J, the lowest closed contour can be as low as $A_\mathrm{V} \approx 2$ mag. For brevity, we refer to this second cloud definition as a "$A_\mathrm{V} = 7$ mag box" in the remainder of the paper.

\section{Results}           
\label{sec:results}

\subsection{High-dynamic range column density maps of IRDCs}
\label{subsec:hdr_maps}


We used the method described in Section \ref{subsec:combination} to derive column density maps for the ten IRDC complexes of our sample. We present the maps in Appendix \ref{app:data} (online only). The pixels in the maps for which there was either no MIR or NIR data available are blanked to black. These blanked areas can result from strong MIR point sources and nebulosity, or missing \emph{UKIDSS} data (due to the incomplete coverage of the survey).

\begin{figure*}
   \centering
\includegraphics[bb = 0 0 680 680, clip=true, width=0.95\textwidth]{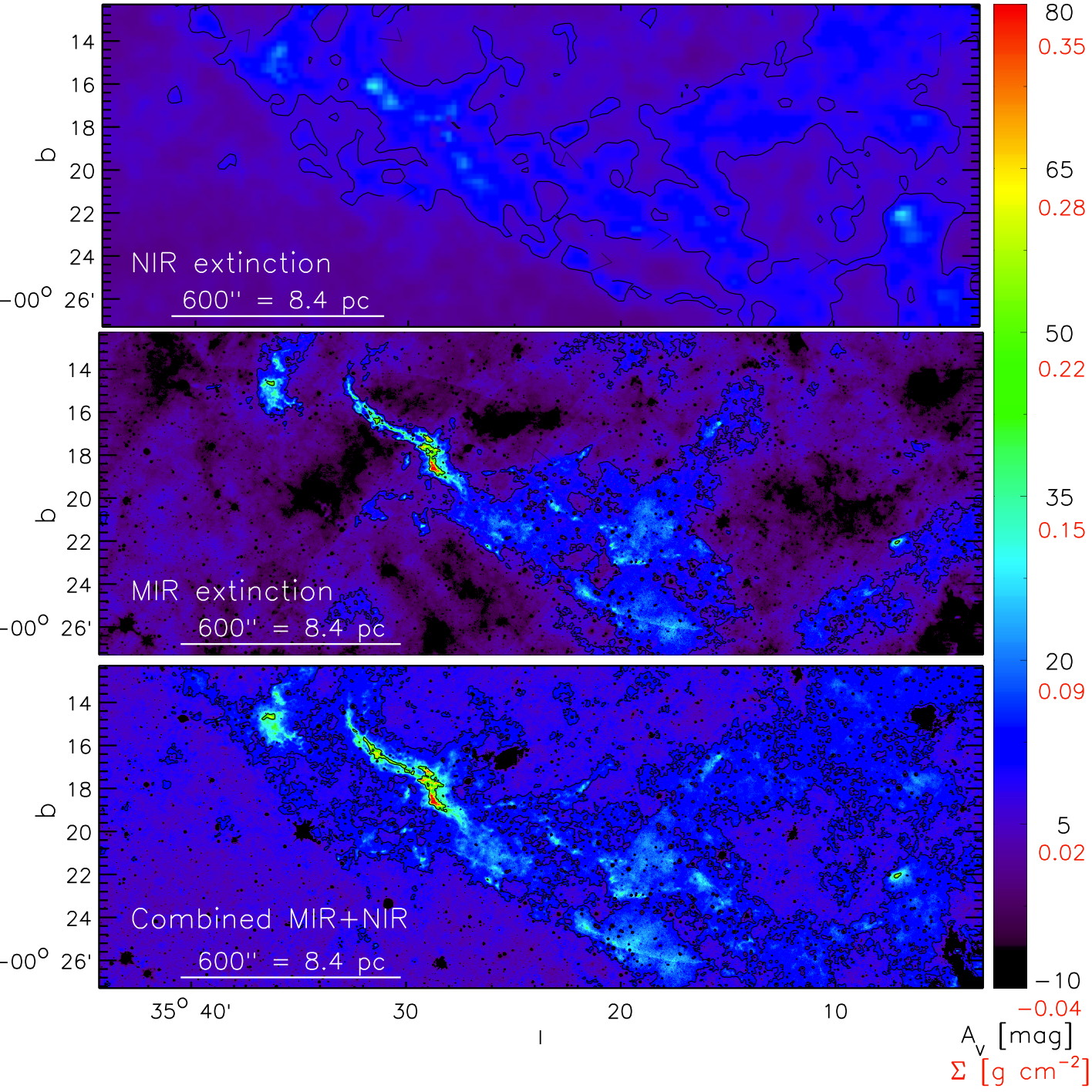}
\includegraphics[bb = 35 55 550 340, clip=true, width=0.65\textwidth]{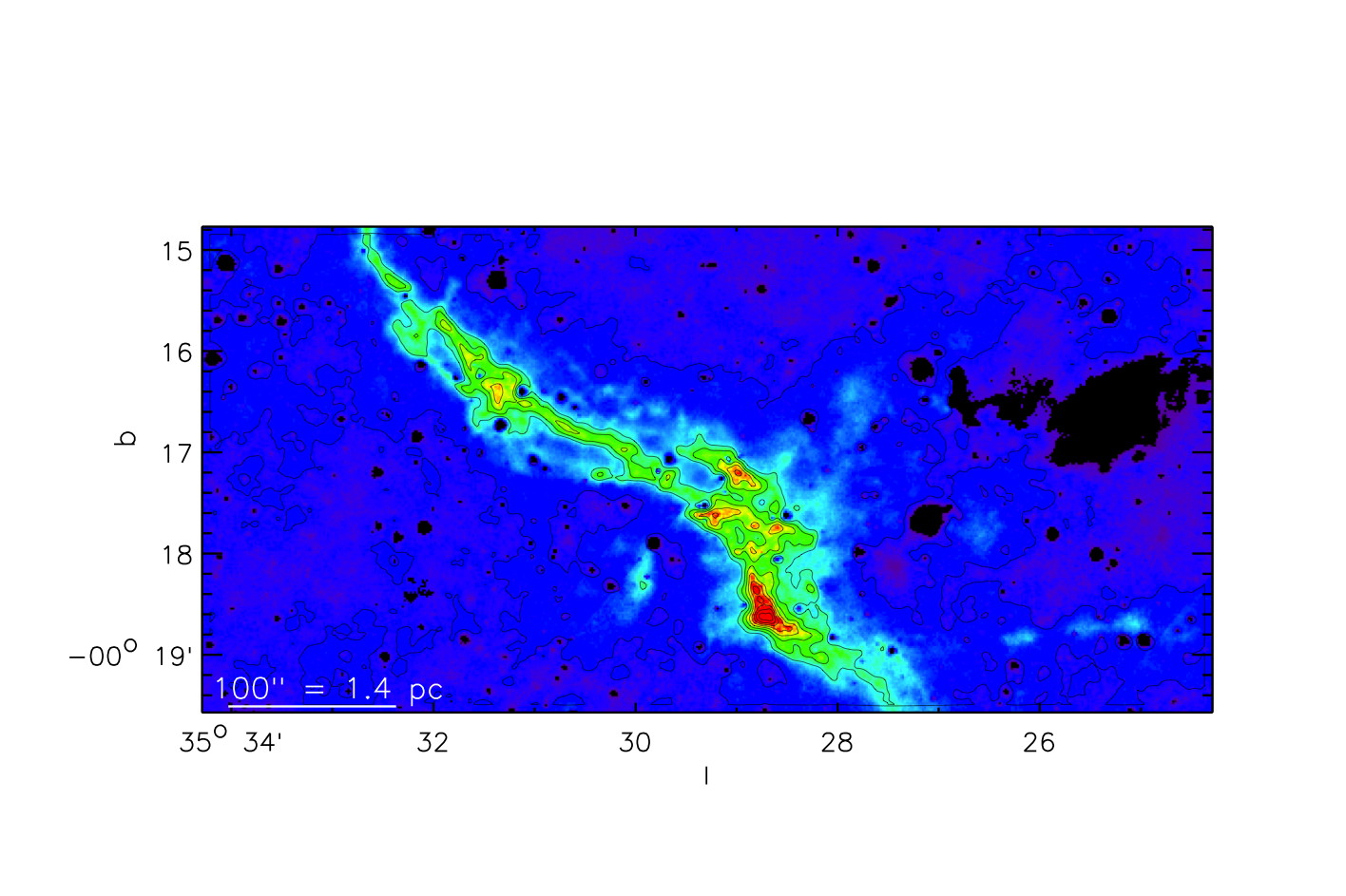}
      \caption{Column density maps of the H cloud. The contours in the three-panel figure are at $A_\mathrm{V} = \{7, 40\}$ mag. The color scale is given in units of visual extinction, $A_\mathrm{V}$ (black numbers), and mass surface density, $\Sigma$ (red numbers). \emph{Top: }The map based only on NIR data \citep{kai11a}. \emph{Middle: }The map based only on 8 um data (BT12). \emph{Bottom: }The map based on combined NIR and MIR data. The standard deviation of column density values in a region about $\sim 10\arcmin$ distance from the cloud where $A_\mathrm{V} < 2$ mag is $\sigma (A_\mathrm{V}) = 1.4$ mag. The maximum values in the map are $A_\mathrm{V} \approx 140$ mag. The separate panel at the bottom shows a blow-up of the combined extinction map. The contours are at $A_\mathrm{V} = \{ 7, n \times 10 \}$ mag.
              }
         \label{fig:showcase}
   \end{figure*}

Figure \ref{fig:showcase} shows as an example the column density map of the IRDC H and, for comparison, the separate NIR- and MIR-derived column density maps of the complex \citep[][BT12, respectively]{kai11a} in the same color-scale. The combined map clearly reflects the characteristics of both data sets: it recovers the large-scale structures in the cloud with better sensitivity than the MIR data alone, and it retains the high angular resolution of high-column density structures of the original MIR-derived maps. The combined maps probe column densities from the sensitivity limit ($\sigma (A_\mathrm{V}) \approx$ 1 mag) to the saturation limit of the MIR data ($A_\mathrm{V, sat} \simeq$ 100 mag).

We list in Table \ref{tab:clouds} the new masses of the IRDCs calculated from the combined column density data. The masses from the combined data are on average 1.4 times larger than the masses calculated from the MIR data alone. We also show in Table \ref{tab:clouds} also the virial parameters of the clouds, defined as \citep{ber92}
\begin{equation}
\alpha_\mathrm{vir} = \frac{5 \sigma_\mathrm{v}^2 R}{GM}.
\end{equation}
In this $\sigma_\mathrm{v}$ is the velocity dispersion in km s$^{-1}$ for which we use the dispersions derived using Gaussian fits and ellipsoidal cloud definition (i.e., $\sigma_\mathrm{G, ell}$ in Table \ref{tab:clouds}). $R$ is the radius of the ellipsoid in pc for which we use the expression $R = \sqrt{\mathrm{area} / \pi}$, and $M$ is the mass in solar masses. $G = 1/232$ is the gravitational constant in corresponding units. Five out of ten clouds have their virial parameters within the factor of 2 of unity. The cloud G has the largest virial parameter, $\alpha_\mathrm{vir} = 4.7$. None of the clouds have their virial parameters more than factor of 2 below unity. Modulo the effects of surface pressure terms, internal density structure, and cloud elongation, these results suggest that IRDCs (at least those typical of this sample) tend to be self-gravitating, gravitationally bound objects \citep[see also][]{her12}.


\subsection{Probability distributions of column densities}
\label{subsec:pdfs}


We used the derived maps to examine the column density PDFs in the IRDCs. The PDFs describe the probability of observing a column density value $A_\mathrm{V}$ in the clouds. Because of the completeness issues in defining the clouds with the \citet{sim06} ellipsoids, we only examine the PDFs within the $A_\mathrm{V} = 7$ mag boxes (see Section \ref{subsec:area}), and analyze them only above $A_\mathrm{V} \ge 7$ mag. We also note that we do not use all the dynamic range present in the column density maps which extend typically to $A_\mathrm{V} = 140-150$ mag. As discussed by BT12 the saturation of the mid-infrared technique starts to affect the column density values clearly earlier. The saturation limits, $\Sigma_\mathrm{sat}$, below which BT12 conclude the maps to be relatively unaffected by saturation, are given in Table \ref{tab:clouds}. The error in column density due to saturation is for most clouds $\lesssim 10\%$ when $\Sigma \lesssim 0.8 \Sigma_\mathrm{sat}$. We only include the column densities up to the saturation limits $\Sigma_\mathrm{sat}$ to the PDFs.

\begin{figure}
   \centering
\includegraphics[width=\columnwidth]{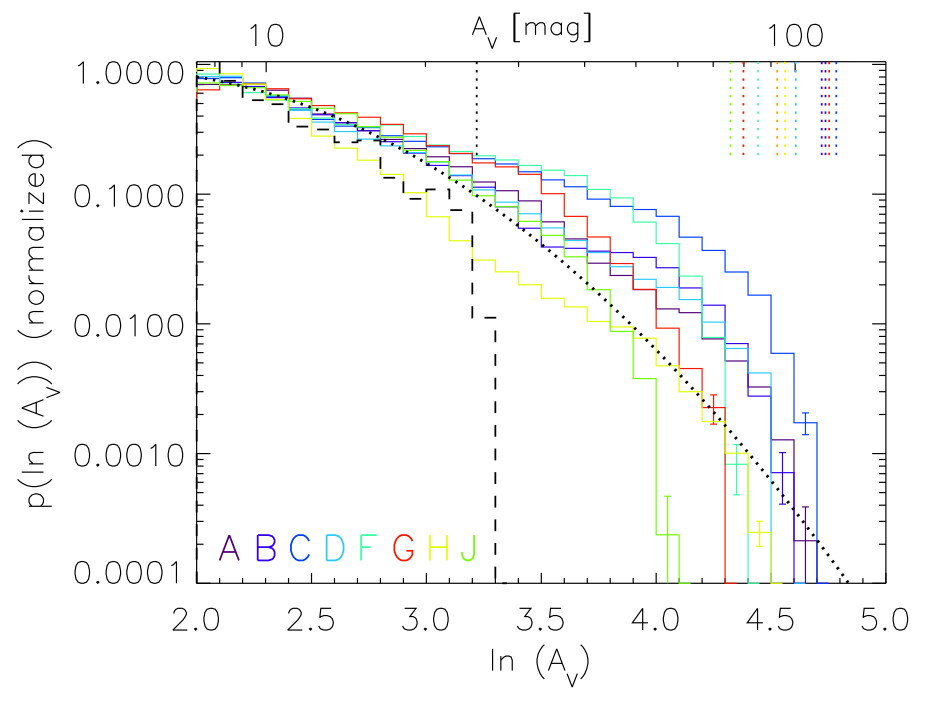}   
      \caption{Column density PDFs of the IRDCs in the sample. The PDFs are normalized so that a linear fit to the range $\ln{A_\mathrm{V}} = 2-3$ fulfills the boundary condition $p(\ln{A_\mathrm{V}}= 2.0) = 1$. The dashed histogram shows the PDF of the Ophiuchus cloud from K09. The black dotted vertical line marks the upper limit of the K09 data ($A_\mathrm{V} = 25$ mag). The black dotted curve shows, for reference, a log-normal function with $\mu = 1.2$, $\sigma_\mathrm{\ln{N}} = 0.9$ (not a fit). The colored vertical dotted lines show the saturation limits of the IRDCs.
              }
         \label{fig:pdfs}
   \end{figure}

Figure \ref{fig:pdfs} shows the PDFs of eight IRDCs of our sample. The PDFs were normalized for the plot by first fitting a line to the range $\ln{A_\mathrm{V}} = 2-3$ ($A_\mathrm{V} = 7.4-20$ mag) and then setting a boundary condition $p(\ln{A_\mathrm{V}}= 2.0) = 1$. In general, the PDFs of all IRDCs are very similar to each other below $A_\mathrm{V} \lesssim 30-40$ mag. Above that, they differ from each other, with the PDFs of some clouds decreasing more rapidly than others (i.e., they contain a lower fraction of high column densities). For reference, we show in Fig. \ref{fig:pdfs} a log-normal function with $\mu = 1.2$, $\sigma_{\ln{N}} = 0.9$ (see Eq. \ref{eq:lognormal}). The PDFs of all but two IRDCs (C and D) are within a factor of three of this curve throughout their dynamic range. The individual clouds show varying behaviors, e.g., the cloud C has almost a power-law like PDF until $A_\mathrm{V} \approx 70$ mag, after which it drops abruptly. It is not easy to assess whether the PDFs are better fit with a power-law between $\ln{A_\mathrm{V}} \approx 2-3.5$ and a drop afterwards, or by a log-normal. This is because of the similarity of the power-law and a wide log-normal function whose peak (mode) is not covered. We analyze the properties of the PDFs further in Sections \ref{subsec:moments}-\ref{subsec:disc_var}.

We also note as a caveat that the pixels affected by bright MIR sources (YSOs) and nebulae are blanked in the extinction data and can affect the PDFs (see Section \ref{subsec:combination}). In principle, it is possible that especially the high-extinction portions of the PDFs are affected by the missing data, because high-column density regions could presumably harbor more active star formation than low-column density regions. There are bright MIR point sources, likely local to the cloud, present inside the \citet{sim06} ellipsoid areas of the clouds A, C, D, E, and H. However, the area covered by these blanked regions is only a fraction of, e.g., the area above $A_\mathrm{V} > 40$ mag, which gives some suggestion that their effect should not be very large. Similarly, while almost every cloud has some MIR-bright nebulosity in their area, only in the case of Cloud I is the fraction of blanked pixels higher than a few percent (for the cloud I about 10\% of the "7 mag box" is affected, see Section \ref{subsec:area}). Therefore, we note that some caution is in place when regarding the PDF of the Cloud I, and we also exclude it from the analyses of the PDFs performed later in this paper.


\subsection{Fractional amount of high-column density material}
\label{subsec:cpdfs}

\begin{figure}
   \centering
\includegraphics[bb = 20 10 320 340, clip=true, width=0.5\columnwidth]{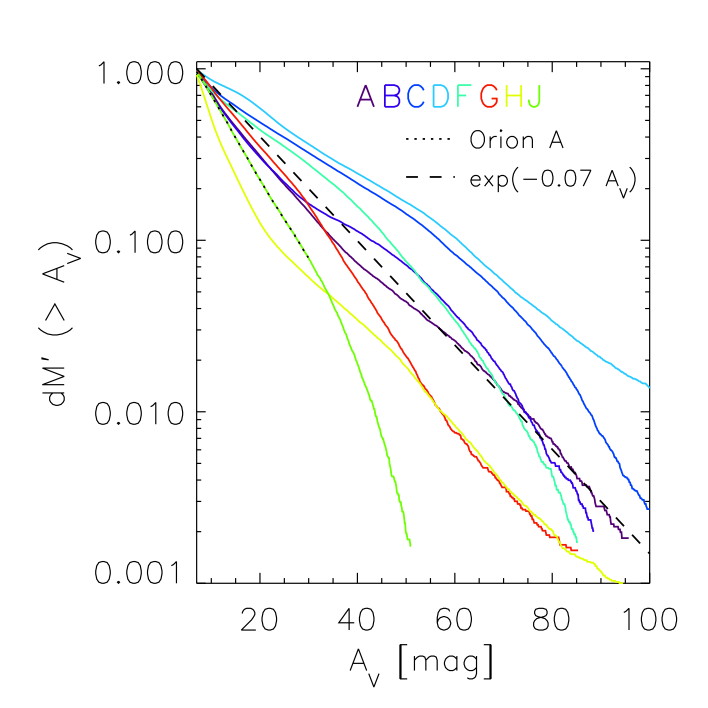}\includegraphics[bb = 20 10 320 340, clip=true, width=0.5\columnwidth]{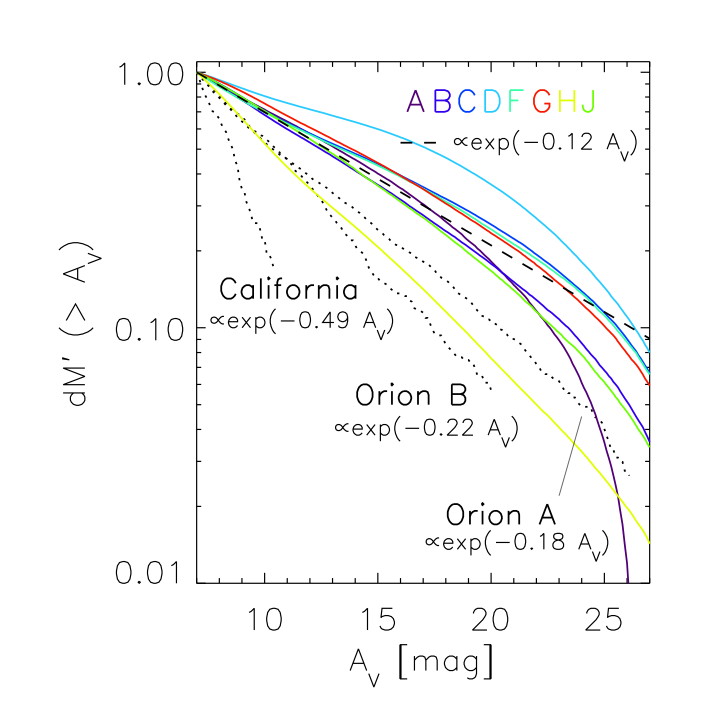}
      \caption{\emph{Left: }Cumulative distribution functions (CDFs) of the IRDCs, expressing the fraction of mass at column densities higher than the abscissa value. The functions are normalized to unity at $A_\mathrm{V} = 7$ mag. For comparison, the CDF of the Orion A molecular cloud from K09 is shown in the panel. The functions are normalized at their value at $A_\mathrm{V} = 7$ mag. \emph{Right: }The CDFs of the IRDCs and three nearby clouds from K09 (Orion A, Orion B, and the California Cloud, shown with dotted lines). The CDFs of the IRDCs were derived after smoothing and re-binning the IRDC column density data to the spatial resolution of the nearby clouds. Only the values in the range $A_\mathrm{V} = 7-30$ mag were used in deriving the CDFs shown in the panel.
              }
         \label{fig:cumpdfs}
   \end{figure}

We examined the fractional amount of high-column density material in the IRDCs by calculating the cumulative distribution functions (CDFs) of their column densities. Here we define the CDF as the fraction of mass above the reference value
\begin{equation}
\mathrm{d}M' (> A_\mathrm{V}) = \frac{M(> A_\mathrm{V})}{M_\mathrm{TOT}},
\end{equation}
where $M(> A_\mathrm{V})$ is the mass above the threshold $A_\mathrm{V}$ and $M_\mathrm{TOT}$ is the total mass above $A_\mathrm{V} = 7$ mag.
The CDFs are constructed from the same pixels as the PDFs in Section \ref{subsec:pdfs}, i.e., for the $A_\mathrm{V} = 7$ mag box. The resulting CDFs are shown in Fig. \ref{fig:cumpdfs} (left panel). The CDFs show a relatively linear shape in the log-linear scale, i.e., their shapes are consistent with exponential functions. We find that on average, the CDFs of the clouds follow the form
\begin{equation}
\mathrm{d}M' (> A_\mathrm{V}) \propto e^{ -0.07 \times A_\mathrm{V} }.
\label{eq:cmf_exp}
\end{equation}
This relation was obtained through a fit to the mean CDF of all IRDCs. However, we note that the CDF, in general, depends on the dynamic range and resolution covered by the observations. 
Remembering this, we compared the CDFs derived for IRDCs with those of nearby molecular clouds from K09. To facilitate the comparison, the dynamic ranges of the IRDC data and the data from K09 were truncated to the range covered by both data sets. The lower limit for this range is set by the IRDC data, which is complete only above $A_\mathrm{V} \gtrsim 7$ mag. The upper limit is set by the data of K09 that probe only values up to about $A_\mathrm{V} \lesssim 30$ mag. Therefore, we constructed the CDFs for this comparison using only the range $A_\mathrm{V} = 7-30$ mag.

Figure \ref{fig:cumpdfs} (right panel) shows the resulting CDFs for the IRDCs and three nearby giant molecular clouds (Orion A, Orion B, and the California Cloud). These nearby clouds have similar masses ($M \approx 10^5$ M$_\odot$), but their star-forming activities differ significantly. Orion A is the most active cloud with 2862 member sources, Orion B has 635 members, and the California Cloud only 279 members \citep[adopting the member counts from][]{lad09}. Fits of Eq. \ref{eq:cmf_exp} to the CDFs of these clouds result in slopes -0.18, -0.22, and -0.49 for the Orion A, Orion B, and California Cloud, respectively. The average CDF of the IRDCs has an exponent of $-0.12$, which is even flatter than that of Orion A. The main source of uncertainty in the CDF of the IRDCs originates from the chosen absolute value of $\kappa_8$ (see Section \ref{subsec:combination}). To first-order, the uncertainty in $\kappa_8$ reflects as a multiplier to the CDF: for example, underestimation of $\kappa_8$ by a factor of 2 results in over-estimation of column densities by a factor of roughly two, and hence, the exponent of the CDF will be under-estimated by the same factor. In Section \ref{subsec:mirdata} we discussed that the absolute uncertainty of $\kappa_8$ is $\sim 30$\%. This translates to uncertainty of $\sim 30$\% for the exponent of the CDF. Hence, the exponent -0.12 we derive from the IRDCs is roughly 2-$\sigma$ higher than that of Orion A. We conclude that this implies that the fraction of high-column density material in IRDCs is at least comparable, and possibly higher than in nearby active star-forming clouds. This result is similar to what we found out in \citet{kai11a} using only NIR data. With the data presented in this paper we quantify the form of the CDF above $A_\mathrm{V} > 7$ mag and establish that it on average has an exponential shape that continues up to $A_\mathrm{V} \lesssim 100$ mag.


\section{Discussion}     
\label{sec:discussion}

\subsection{Column density probability distributions of IRDCs}  
\label{subsec:disc_pdfs}


How do the PDFs we derive for IRDCs in this paper compare with earlier findings of the column density PDFs in molecular clouds? We show in Fig. \ref{fig:pdfs} the PDFs of the IRDCs and that of the Ophiuchus cloud from K09, which is representative of most star-forming clouds. The PDF of Ophiuchus is very similar to those of the IRDCs, and consistent with either a power-law or a log-normal ($\mu = 1.2$, $\sigma_\mathrm{\ln{N}} = 0.9$). Thus, the PDFs of IRDCs seem to resemble the PDFs of nearby star-forming clouds, although they are slightly shallower, i.e., they contain relatively more high-column density material.

\begin{figure}
   \centering
\includegraphics[bb = 0 10 590 400, clip=true, width=\columnwidth]{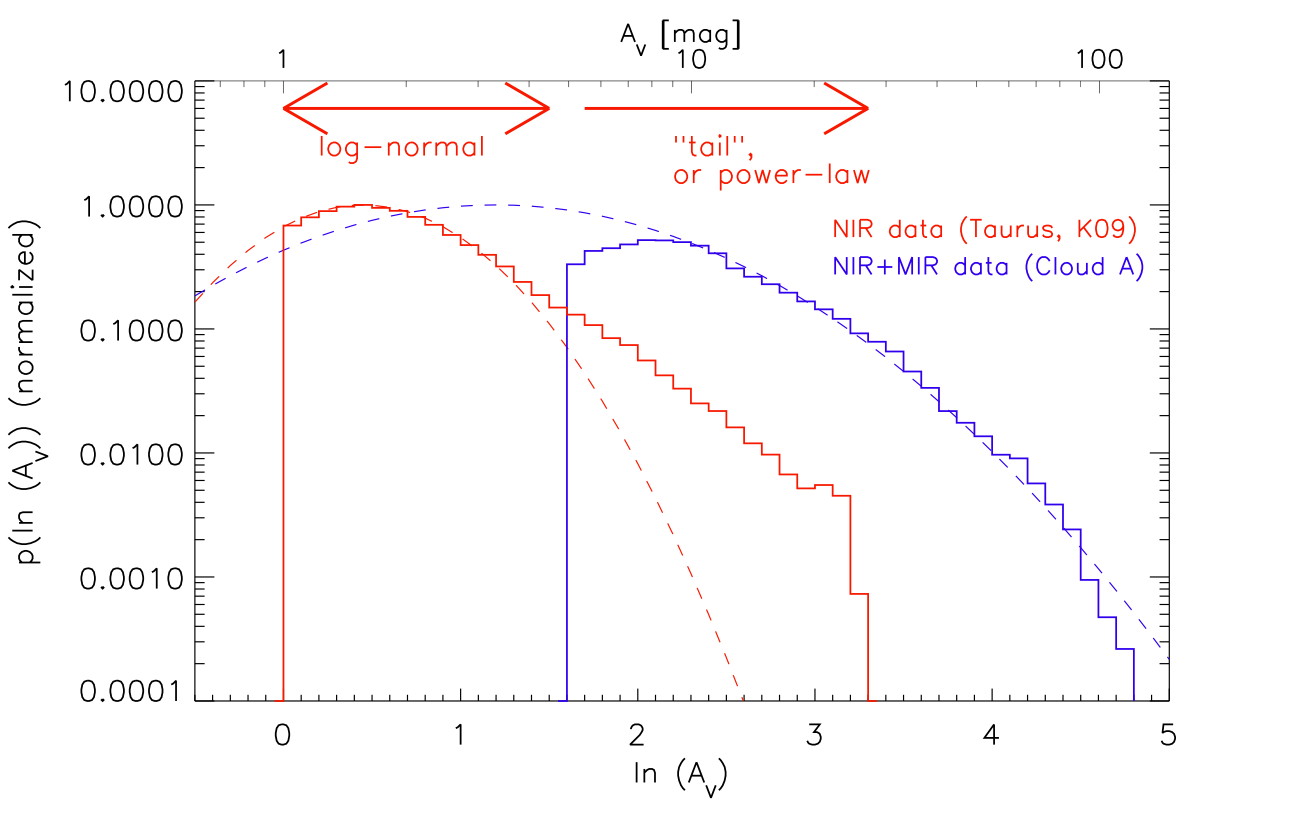}
      \caption{Column density PDFs of a nearby star-forming cloud (Taurus, red histogram) from K09 and one IRDC of this study (Cloud A, blue histogram). In K09, the different parts of the PDFs are referred to as "log-normal" and a "tail" (or, a "power-law)". The ranges of these two components are indicated in the figure with red arrows. The red dashed line shows a log-normal fit to the "log-normal" part of the PDF. The blue dashed line that shows a log-normal fit to the IRDC data results to a clearly \emph{wider} log-normal function than those discussed by K09. As the figure indicates, the log-normals we fit to IRDC data are relatively close in their shape to the "power-law tails" discussed by K09 in the range $A_\mathrm{V} = 7-25$ mag.
              }
         \label{fig:cartoon}
   \end{figure}

We note that in K09 we refer to the high-column density tails (at $A_\mathrm{V} \gtrsim 2-5$ mag) of the PDFs as "power-law" tails. We elucidate the nomenclature and the dynamic ranges of the K09 data and this work in Fig. \ref{fig:cartoon}. The power-law-like behavior of the PDF of Ophiuchus is obvious in Fig. \ref{fig:pdfs}. However, the PDF of Ophiuchus is, as well as the PDFs of the IRDCs are, consistent with the log-normal function shown in the figure. This log-normal function is significantly \emph{wider} than those used by K09 to describe the PDFs of low-column density ($A_\mathrm{V} \lesssim 2-5$ mag) material. Thus, the data we present in this paper suggests that also the high column densities in molecular clouds can be described by a log-normal PDF, however, one clearly wider than the log-normal that we used in K09 to describe low column densities. It remains to be explored in future studies whether the column density PDFs of IRDCs are well-described by log-normals down to low column densities ($A_\mathrm{V} \approx 1$ mag), or whether they show a similar change in shape at around $A_\mathrm{V} = 2-5$ mag as the nearby star-forming clouds do.

In K09 we showed that the non-star-forming clouds do not show similar high-column density tails as the star-forming clouds do (or, considering the discussion above, their PDFs are clearly narrower than those of the star-forming clouds). This could be interpreted as gravity having a significant effect in shaping the cloud structure as has been established in numerical studies \citep[e.g.,][]{kle00, bal11}, but also possibly as confinement of non-self-gravitating clumps by the pressure imposed to them by their surrounding medium \citep{kai11b}. What is common in both of these frameworks is that the shape of the PDFs, and especially the presence of a wider, log-normal (or power-law-like) component, is intimately linked to the onset of star formation. Further, \citet[][]{row11} analyzed the PDF shapes and embedded populations of nearby clouds and suggested that the PDF shape is also linked to the star-formation mode (isolated/dispersed vs. clustered) of the clouds. The PDFs of the IRDCs  resemble more closely the PDFs of the star-forming clouds than non-star-forming clouds (cf., Figs. \ref{fig:pdfs}, \ref{fig:cumpdfs}) in a sense that the fraction of high-column densities is high (regardless whether the PDF shape is log-normal or power-law). This alone suggests that substantial (low-mass) populations may be present in the IRDCs of our sample. 


We established in Section \ref{subsec:cpdfs} that the fractional amount of high-column density material in IRDCs in the range $A_\mathrm{V} = 7-30$ mag at least comparable to, and possibly even higher than in active nearby clouds. While there is a well-known correlation between the gas surface density and star-forming rate \citep[SFR; ][]{sch59, ken98}, it was recently proposed by \citet{lad12} that there exists a more fundamental scaling between the surface density of \emph{dense} gas and SFR. In particular, \citet{lad12} suggested that the surface density of SFR depends linearly on the mass surface density of H$_\mathrm{2}$
\begin{equation}
\Sigma_\mathrm{SFR} \propto f_\mathrm{DG} \Sigma (\mathrm{H}_2),
\end{equation}
where $f_\mathrm{DG}$ is the fraction of dense gas in the clouds, defined as
\begin{equation}
f_\mathrm{DG} = \frac{M(A_\mathrm{K} > 0.8 \mathrm{\ mag})}{M(A_\mathrm{K} > 0.1 \mathrm{\ mag})}.
\end{equation}
Here, $A_\mathrm{K}$ is the extinction in the $K$ band ($A_\mathrm{K} = 1$ mag is equivalent to $A_\mathrm{V} \approx 8.8$ mag, Eq. \ref{eq:cardelli}).
The fraction $f_\mathrm{DG}$ was measured and analyzed in a number of nearby clouds by \citet{lad10}, who derived the mean value $\langle f_\mathrm{DG} \rangle = 0.1\pm 0.06$, with maximum fractions of $f_\mathrm{DG} \approx 0.2$ in the Orion A and Corona Australis clouds. Because the PDFs of the IRDCs we present in this paper are complete only above $A_\mathrm{V} > 7$ mag, we cannot directly calculate the values of $f_\mathrm{DG}$ from our data. However, we can present an estimate for the $f_\mathrm{DG}$ in the following manner. We re-examined the CDF data presented by K09. We fitted exponential functions to the high-column density slopes of the CDFs of all star-forming clouds (13 clouds) and extrapolated the clouds' masses above $A_\mathrm{K} > 0.1$ mag using the fitted functions. This test showed that the cloud masses were underestimated on average by a factor of $ \langle c \rangle = 1.9^{+0.8}_{-0.5}$ using this procedure. The scatter in the factor had the 3-$\sigma$ interval of $c=[0.7, 5.2]$. 

Since the cloud masses are systematically underestimated, we applied the factor $\langle c \rangle = 1.9$ to the masses of IRDCs derived by extrapolating their exponential CDFs to lower column densities. This resulted to the average dense gas mass fraction of $f_\mathrm{DG} \approx 0.2$ of our IRDC sample, which is about two times larger than the average fraction derived for nearby clouds by \citet{lad10}. Given the uncertainty of our analysis, we find that this implies that the $f_\mathrm{DG}$ in IRDCs is of similar level, and possibly even higher, than in nearby clouds. This, in turn, suggests that the SFR of the IRDCs should be at least comparable to equally massive nearby molecular clouds on average. We recognize that this prediction remains suggestive, because we could not directly calculate the dense gas mass fraction, but had to rely on a correction for the missing mass below $A_\mathrm{V} \lesssim 7$ mag.

\subsection{Relationship between the sonic Mach number and column density variance in molecular clouds}
\label{subsec:var_intro}

We demonstrate in this section the potential of high-dynamic-range column density data to examine the relationship between the column density dispersion and the sonic Mach number in molecular clouds. This relation is interesting, because it is directly linked to the fundamental relation between density fluctuations and turbulent energy in supersonically turbulent media: schematically, higher turbulent energy (higher Mach number) is expected to induce stronger density fluctuations, and hence, \emph{wider} density PDF. In the following, we describe the formalism of this connection in more detail. 

Numerical simulations predict a log-normal form for the PDF of volume densities in non-gravitating, isothermal media \citep[e.g.][]{vaz94, pad97mnras}.
The dispersion of the logarithmic standard deviation $\sigma_\mathrm{\ln{x}}$ (see Eq. \ref{eq:lognormal}) is linked to the turbulent energy, i.e., the (non-thermal) sonic Mach number, ${\cal M}_\mathrm{s}$, via the equation \citep[e.g.,][]{nor99, ost99}
\begin{equation}
\sigma_\mathrm{\ln{x}}^2 = \ln{(1+{\cal M}_\mathrm{s}^2 b^2)},
\end{equation}
where $b$ is a proportionality constant. This equation translates to the dispersion of non-logarithmic, mean-normalized volume densities
\begin{equation}
\sigma_{\rho /  \langle \rho \rangle} = b {\cal M}_\mathrm{s}.
\label{eq:b}
\end{equation}
Thus, the proportionality constant $b$ concretely describes how strong is the coupling between the turbulent energy and density fluctuations. Exploring the value of $b$ in numerical simulations  \citep[e.g.,][]{pad97mnras, kri07, fed08b, lem08, pri11} and analytically \citep[e.g.,][]{mol12} has received ample attention recently, owing to its intimate connection to the analytical formulations of the dense core formation rate and the possible origin of the IMF \citep[e.g.,][]{pad02, kru05, hen08, elm11}.

While observations of column density only probe the 2D density statistics, it is established that the log-normal nature of the volume density PDF remains in 2D to 3D transformation \citep[e.g.,][]{vaz01, fed10}. Taking this statement one step further, \citet{bru10method} presented a formalism to directly estimate the 3D density variance from the measurements of 2D column density field. Their method is based on the possibility to estimate the 3D-2D variance ratio
\begin{equation}
R = \frac{ \sigma^2_{N / \langle N \rangle} }{\sigma^2_{\rho / \langle \rho \rangle}}, 
\label{eq:R}
\end{equation}
using solely observations of 2D density field. In the equation $\sigma_{N / \langle N \rangle}$ is the standard deviation of mean-normalized \emph{column} densities. We note that \citet{bru10method} examined the values of $R$ in a set of isothermal simulations of supersonic turbulence and found that in all their models $R$ was between $R = [0.03, 0.15]$.
 If we now combine Eq. \ref{eq:R} with Eq. \ref{eq:b}, we have an expression for $b$ which depends only on 2D statistics
\begin{equation}
b = \frac{\sigma_{\rho / \langle \rho \rangle}}{\cal M_\mathrm{s}} = \frac{\sigma_{N / \langle N \rangle}}{\cal M_\mathrm{s}} R^{-\frac{1}{2}} = a_1 \times R^{-\frac{1}{2}}.
\label{eq:slope}
\end{equation}
In this equation, $a_1$ is the ratio between the sonic Mach number and column density variance, i.e., the slope potentially measurable (however, see the discussion below) through a linear fit to the $[{\cal M}_\mathrm{s}, \sigma_{N / \langle N \rangle}]$ data
\begin{equation}
\sigma_{N / \langle N \rangle} = a_1 \times {\cal M}_s + a_2.
\label{eq:fit}
\end{equation}
It remains to be explored in future work whether the quality of our combined NIR+MIR data is suitable to directly estimate the variance ratio $R$, and from therein, the 3D variance in IRDCs. Nevertheless even in the absence of the transformation, relationship between the 2D density variance with the sonic Mach number is a proxy of the relationship in 3D. Supporting this approach, recently \citet{bur12} showed that in MHD simulations the ${\cal M}_\mathrm{s} - \sigma_\mathrm{\ln{x}}$ relation is preserved in transformation from 3D to 2D variables, with the details of the correlation depending somewhat on the simulation parameters.


One should emphasize that interpreting observational $({\cal M_\mathrm{s}}, \sigma_{N / \langle N \rangle})$ data in terms of the formalism presented above assumes that the measurements of the column density variance and the mean column density are performed \emph{over such dynamic range that they represent the true underlying total variance and mean}. This is clearly not the case if the column densities are measured only over a relatively narrow dynamic range, as would be the case with, e.g., $^{13}$CO line emission observations \citep[$A_\mathrm{V} \approx 3-6$ mag, e.g.,][]{goo09}.

With this background in mind, we present in the following two approaches for analyzing the ${\cal M}_\mathrm{s} - \sigma_{N / \langle N \rangle}$ relation. First, we describe a calculation in which we simply correlate the column density dispersions measured from the $A_\mathrm{V}$ data above $A_\mathrm{V} \gtrsim 7$ mag with $\cal{M}_\mathrm{s}$. This experiment can serve as a point of comparison for other numerical/observational works, but we cannot directly connect it with the formalism presented earlier in this section. Second, we make an effort to estimate the mean and variance of the observed column density fields in a \emph{dynamic-range-independent manner} by assuming that the underlying column density PDF is log-normal. This assumption allows us to connect the relation directly to Eqs. \ref{eq:b}-\ref{eq:fit} and to estimate how strongly the density fluctuations couple with the turbulent energy, i.e., estimate the coefficient $b$ (Eq. \ref{eq:slope}). 

\subsubsection{Model-independent approach}
\label{subsec:moments}

We first examine the relationship between the sonic Mach number, $\cal{M_\mathrm{s}}$, calculated from the $^{13}$CO line emission and the standard deviation of the mean-normalized column density data, $\sigma^\mathrm{data}_{N / \langle N \rangle}$, defined as\footnote{We use here a notation $\sigma^\mathrm{data}_{N / \langle N \rangle}$ to emphasize that the parameter is calculated directly from the column density data and is \emph{not} necessarily the same as $\sigma_{N / \langle N \rangle}$ discussed in Section \ref{subsec:var_intro}.}
\begin{equation}
\sigma^\mathrm{data}_{N / \langle N \rangle} \equiv \sqrt{ \frac{1}{n} \sum_{i=1}^n \big(\frac{N_i}{\langle N \rangle} - \big< \frac{N_i}{\langle N \rangle} \big> \big)^2},
\label{eq:sigma_N}
\end{equation}
where $n$ is the number of column density values $N$ that enter the calculation and $\langle N \rangle$ is the mean column density in the field. We examined $\cal{M_\mathrm{s}}$ and $\sigma^\mathrm{data}_{N / \langle N \rangle}$ only in the case of the $A_\mathrm{V} = 7$ mag box, because of potential completeness issues in $\sigma^\mathrm{data}_{N / \langle N \rangle}$ with ellipsoidal cloud definition. Only data above $A_\mathrm{V} \ge 7$ mag was considered. The Mach numbers were calculated from the one-dimensional velocity dispersions derived in Section \ref{subsec:co} (given in Table \ref{tab:clouds}) with the expression
\begin{equation}
{\cal M}_\mathrm{s} = \frac{\sigma^{3D}_\mathrm{CO}}{c_\mathrm{s}} =  \frac{\sqrt{3} \sigma^{\mathrm{1D}}_\mathrm{CO}}{c_\mathrm{s}},
\label{eq:Ms}
\end{equation}
where $c_\mathrm{s} = (kT / \mu m_\mathrm{H})^{1/2}$ is the isothermal speed of sound and the velocity dispersion of CO is assumed to represent turbulent velocity. In the lack of measurements of the gas kinetic temperatures, we chose $T=15$ K, consistent with the observations of \citet{car98} and \citet{pil06}, and $\mu = 2.33$ for the calculation of the speed of sound for all IRDCs. Eq. \ref{eq:Ms} assumes that the velocity field is isotropic and the 3D velocity dispersion is related to the line-of-sight velocity dispersion with $\sigma_\mathrm{3D} = \sqrt{3} \sigma_\mathrm{1D}$. 

\begin{figure*}
   \centering
\includegraphics[width=\columnwidth]{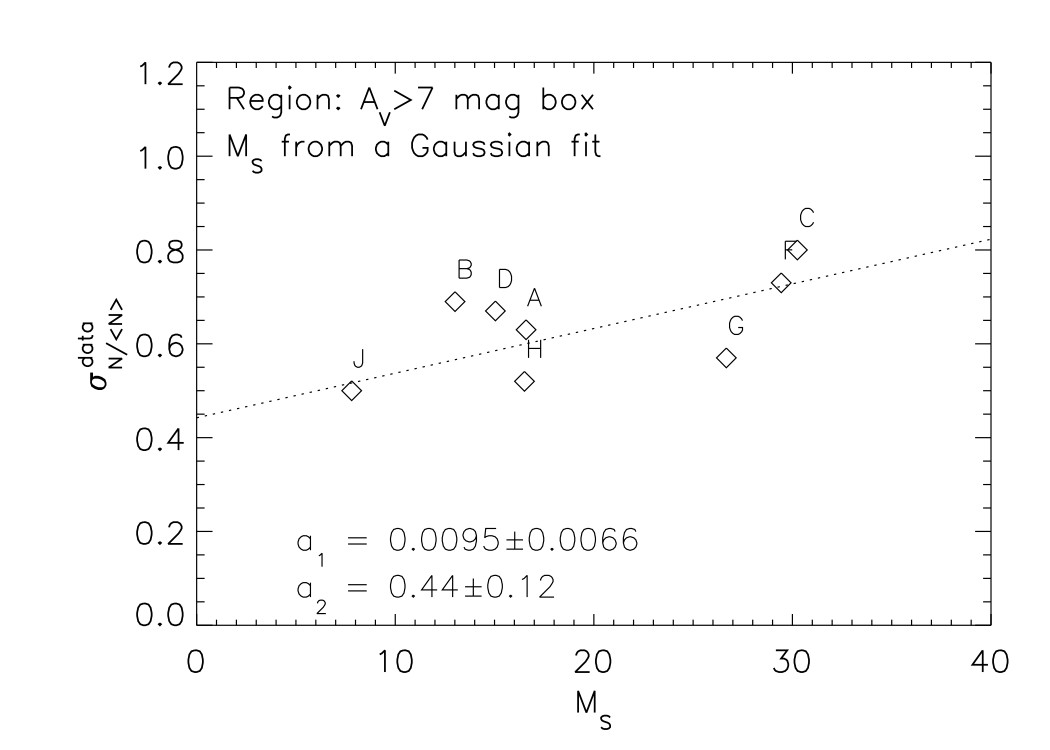}\includegraphics[width=\columnwidth]{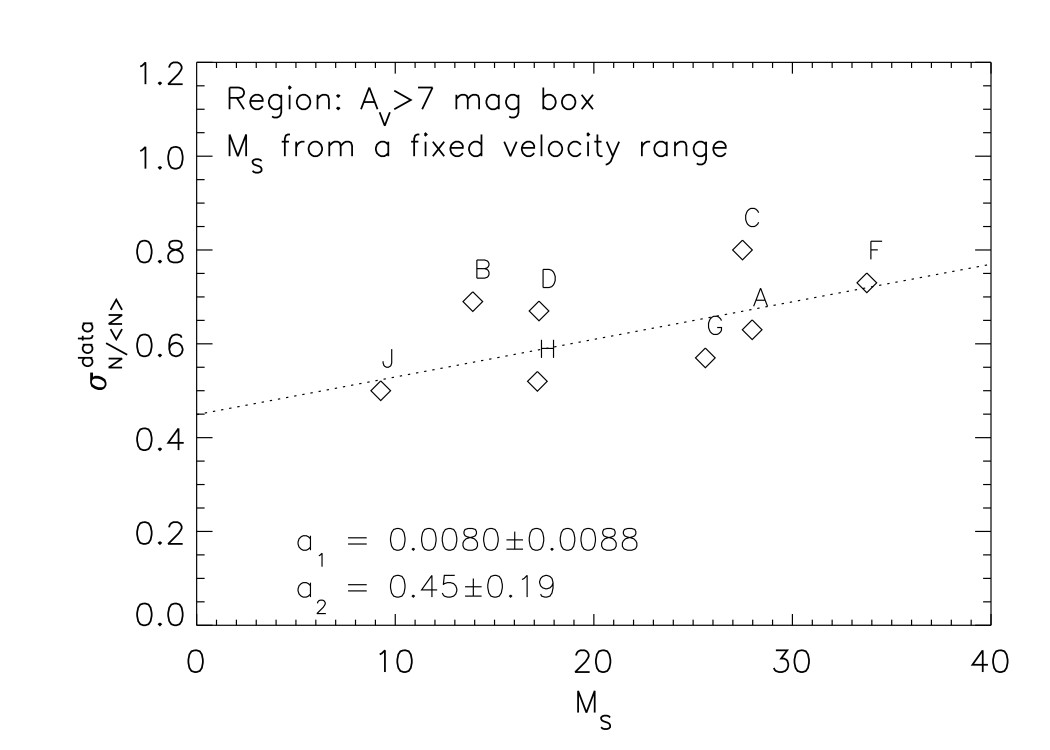}
      \caption{Standard deviation of the mean-normalized column densities, $\sigma^\mathrm{data}_{N / \langle N \rangle}$, as a function of the sonic Mach number, ${\cal M_\mathrm{s}}$, in eight IRDCS. The Cloud E is not included in the plot, because of the potential problems in the extinction data of the cloud (see Section \ref{subsec:co}). The Cloud I is also excluded because of nebulosity in the cloud region. The left frame shows the relation when ${\cal M_\mathrm{s}}$ is calculated from Gaussian fits to the mean spectra. The right frame shows the relation when ${\cal M_\mathrm{s}}$ is calculated from a fixed velocity range (see Section \ref{subsec:co}). The $A_\mathrm{V} = 7$ mag box was to define the cloud region that was included in the plots (see Section \ref{subsec:area}).
                    }
         \label{fig:var_vs_var}
   \end{figure*}

Figure \ref{fig:var_vs_var} shows the relation between ${\cal M_\mathrm{s}}$ and $\sigma^\mathrm{data}_{N / \langle N \rangle}$ in the IRDCs. The figure shows the data separately  for the two different methods to calculate ${\cal M_\mathrm{s}}$ (see Section \ref{subsec:co}). The Mach numbers of the clouds span the range ${\cal M_\mathrm{s}} = 8-35$, while the column density dispersions extend over a narrower range of $\sigma^\mathrm{data}_{N / \langle N \rangle} = 0.4-0.8$. This range is in agreement with BT09 who concluded that there are differences in the widths of the column density PDFs of these clouds (BT09 used the 'ellipsoid' area in determining their cloud region; see Section \ref{subsec:area}).
The small number of clouds in our sample limits the ability to establish whether the two variables or correlated or not. Also, the uncertainties in both variables are likely dominated by the choices made in defining the cloud region instead of statistical uncertainty. The Pearson's correlation coefficients for both choices of velocity dispersions are below 0.8 (which corresponds to the \emph{p}-value of $p\approx 0.02$ when degrees of freedom is 6). We also fitted the data with the \textsf{linfitex} routine from the \textsf{mpfitfun} library \citep{mar09}. In the fitting, we used as uncertainties the relative uncertainty of 20\% for $\sigma^\mathrm{data}_{N / \langle N \rangle}$ (originating mostly from the zero-point error of the data, $\sigma (A_\mathrm{V}) \approx 1$ mag) and 30\% for ${\cal M_\mathrm{s}}$ (originating mostly from the uncertainty in temperature, which is typically 10-20 K in IRDCs). The resulting slopes (and their errors) imply that the correlation of variables is not significant. However, we emphasize here that, as explained in Section \ref{subsec:var_intro}, this relation cannot be straightforwardly linked to the theory describing the ${\cal M_\mathrm{s}} - \sigma_{\rho / \langle \rho \rangle}$ relation in turbulent media. We also note that uncertainty in opacity-law somewhat increases the uncertainty of the derived relationship (see discussion in Section \ref{subsec:disc_var}).


\subsubsection{Approach based on a log-normal column density PDF} 
\label{subsec:disc_var}


We can connect the observed column density PDFs and $\cal{M}_\mathrm{s}$ values intimately into the theory of turbulent media (Section \ref{subsec:var_intro}) by interpreting the PDFs shown in Fig. \ref{fig:pdfs} as the high-column density sides of log-normal-shaped PDFs and estimating the shape parameters of the underlying PDFs through fits of log-normal functions. The log-normal function is given by Eq. \ref{eq:lognormal}, and we use in it the variable $x = N$, which is the (non-normalized) column density. With this definition, $\mu$ and $\sigma_\mathrm{\ln{N}}$ are then the mean and standard deviation (in logarithmic units) of the log-normal distribution of column densities. Note that this formulation describes the PDF of \emph{non-normalized} column densities: we do not know \emph{a priori} the underlying mean of the column density field which should be used for normalization. From the fit, we determine the parameters $(\mu, \sigma_\mathrm{\ln{N}})$, which are used to calculate the mean, $\langle N \rangle$, and standard deviation, $\sigma_\mathrm{N}$, in non-logarithmic units
\begin{equation}
\langle N \rangle = e^{ \mu } + \sigma_\mathrm{\ln{N}} ^2 / 2,
\end{equation}
\begin{equation}
\sigma_\mathrm{N} = \big[ (e^{ \sigma_\mathrm{\ln{N}}^2 } - 1) e^{ 2 \mu+\sigma_\mathrm{\ln{N}}^2} \big]^{1/2}.
\end{equation}
The standard deviation of the mean-normalized, non-logarithmic column densities is finally the standard deviation of non-normalized data divided by the mean $m$
\begin{equation}
\sigma_{N / \langle N \rangle} = \frac{\sigma_\mathrm{N}}{\langle N \rangle}.
\end{equation}

The parameters resulting from the fit are listed in Table \ref{tab:pdfs}. Figure \ref{fig:var_vs_var2} shows the resulting ${\cal M_\mathrm{s}} - \sigma_{N / \langle N \rangle}$ relation. The Mach numbers were calculated as described earlier in Section \ref{subsec:moments}. We show the relation only for the case when the cloud is defined with the $A_\mathrm{V} = 7$ mag box, because the PDFs of the ellipsoidal definition can suffer from incompleteness (see Section \ref{subsec:area}).

\begin{table}
\caption{Parameters of the log-normal fits to the PDFs}             
\label{tab:pdfs}      
\centering                          
\begin{tabular}{l c c c}        
\hline\hline                 
Cloud 	& $\langle N \rangle$ [mag] 	& $\sigma_\mathrm{N}$ [mag] & $\sigma_{N / \langle N \rangle}$ 	\\    
\hline                        
   A 	&  4.1 	& 5.3 	& 1.3 \\      
   B 	&  6.1	& 4.4 	& 0.7   \\
   C 	&  8.3	& 13.7 	& 1.7 \\
   D 	&  6.7	& 7.5  	& 1.1 \\
   E 	&  -		& - & -		\\ 
   F 	&  4.3	& 8.9 	& 2.1 \\      
   G 	&  5.5	& 6.1 	& 0.9 \\
   H 	&  4.9	& 4.2 	& 0.9\\
   I 	&  -		& - 		& -	\\
   J 	&  8.8	& 6.2 	& 0.7 \\ 
\hline                                   
\end{tabular}
\end{table}


The advantage of using a log-normal model to estimate the shape parameters of the PDF is that allows estimation of the \emph{total} standard deviation of column densities, and from therein, directly connect the shape parameters to the theory of turbulent density fluctuations (see Section \ref{subsec:var_intro}). Under this scheme, we can also combine the ${\cal M_\mathrm{s}} - \sigma_{N / \langle N \rangle}$ measurements of the IRDCs with additional data points from K09 (listed in Table \ref{tab:nearby}, calculated from the data presented in Table 1\footnote{We note that we have subtracted the contribution from noise to the standard deviations reported in K09, which typically is about 0.3 mag. We also note that K09 reports an erroneous for the value of Ophiuchus. Re-analyzing the same data, we get the standard deviation $\sigma \approx 1.9$ mag.} of K09), who measured the standard deviation of column densities in a sample of nearby ($D < 200$ pc) clouds. K09 data allow measurements of the standard deviation of the column densities in the range $1 \mathrm{\ mag}< A_\mathrm{V} \lesssim 25 \mathrm{\ mag}$. Importantly, the \emph{standard deviation measured at this range measures reasonably accurately the standard deviation measured from any wider range, assuming that the shape of the PDFs is even approximately correct}. This is because the probability of having column densities above $A_\mathrm{V} > 25$ mag is, in fact, so small that the contribution of those column densities to the standard deviation becomes negligible. For example, assuming that the true underlying PDF is similar to that of Ophiuchus (see Fig. \ref{fig:pdfs}), but extends onwards from $A_\mathrm{V} > 25$ mag in a power-law (or log-normal) fashion, the standard deviation calculated from the range $1 \mathrm{\ mag}< A_\mathrm{V} \lesssim 25 \mathrm{\ mag}$ is only 5\% smaller than the standard deviation calculated from the all data above $A_\mathrm{V} > 1$ mag. In summary, we can consider the K09 data compatible with the measurements of $\sigma_\mathrm{N / \langle N \rangle}$, because the \emph{total} standard deviation seems to be reasonably well recovered by them. However, we acknowledge that the measurement is based on different approach than the one we use in the case of IRDCs. It remains to be explored with a larger sample of IRDCs which exactly are the conditions under which the assumption of the compatibility is valid.

\begin{table}
\caption{Sonic Mach numbers derived for nearby clouds}             
\label{tab:nearby}      
\centering                          
\begin{tabular}{l c c}        
\hline\hline                 
Cloud 	& ${\cal M}_s$ & $\sigma^\mathrm{data}_{N / \langle N \rangle}$	\\    
\hline                        
   Ophiuchus 	& 10 		& 1.1\\      
   Taurus 		& 10 		& 0.9\\
   Cha I 		& 8.7		& 1.0  \\
   Cha II 		& 12 		& 1.2	\\
   CrA Cloud 	& 6.8 	& 1.1\\ 
   LDN1228 	& 11  	& 0.8\\      
   LDN204 	& 9.6  	& 0.7\\
   LDN1333 	& 14.4 	& 0.6\\
   LDN1719 	& 8.4 	& 1.1\\
   Cha III 		& 11.5  	& 0.8\\ 
   Orion A		& 21		& 1.5      \\
   Per Cloud	& 11		&  1.0     \\
   Orion B		& 17		& 1.2      \\
   Cepheus A	& 22		& 1.1      \\
   California	& 18		&  0.7     \\
\hline                                   
\end{tabular}
\end{table}

 We show in Fig. \ref{fig:var_vs_var2} the ${\cal M_\mathrm{s}} - \sigma_{N / \langle N \rangle}$ values for 14 nearby clouds with red triangles, calculated using the K09 PDF data and \citet{dam01} CO data. We list in Table \ref{tab:nearby} the Mach numbers derived for nearby clouds that span the range ${\cal M_\mathrm{s}} \approx 7-22$. The temperature of $T = 10$ K was adopted in the derivation. We also show in the plot the values derived by \citet{bru10taurus} and \citet{pad97apj} that fall well into the range of our measurements. We note that the velocity dispersions for nearby clouds are derived from $^{12}$CO data that generally probes slightly larger scales than $^{13}$CO data. However the effect of this is relatively small; we calculated that in Perseus, Ophiuchus and Taurus the mean difference between the $^{12}$CO and $^{13}$CO linewidths is a factor of 1.2. We made an experiment in which we multiplied the linewidths of IRDCs by this factor and it does not significantly affect the relations derived in this section.


Again, judging whether the ${\cal M_\mathrm{s}} - \sigma_{N / \langle N \rangle}$ data of our IRDC sample are correlated or not suffers from the low number of clouds. If we treat the measurements for IRDCs and nearby clouds as equals (i.e., assume that both approaches trace reasonably well the \emph{total} column density fluctuations), the Pearson's correlation coefficients for the data are 0.61 and 0.66 for the Gaussian and fixed range velocity dispersions. This coefficient indicates correlation of variables with high significance (degrees of freedom = 20, $p \lesssim 0.001$). We also fitted a linear model to the data again using the \textsf{linfitex} procedure. In this calculation, we used the relative errors of 30\% for the ${\cal M_\mathrm{s}}$ values, and 30\% for the $\sigma_{N / \langle N \rangle}$ values (the error of  $\sigma_{N / \langle N \rangle}$ originates almost entirely from the uncertainty in the mean $\mu$. We use the typical value given by the \textsf{linfitex} procedure). The slopes resulting from this were $a_1 = 0.051 \pm 0.016$ and $a_1 = 0.042 \pm 0.013$ for the Gaussian and fixed range velocity dispersions, respectively. Thus, the derived slopes are approximately 3-$\sigma$ offset from zero. The error-weighted mean of the slopes is $a_1 = 0.047$. We note that choosing the temperature of $T = 15$ K for nearby clouds (uniformly with the value chosen for IRDCs) instead of $T = 10$ K does not affect the significance of the correlation and results to similar slopes ($a_1 = 0.050 \pm 0.015$ and $a_1 = 0.039 \pm 0.012$ for the Gaussian and fixed cases). 

We note that the uncertainty of our adopted $\kappa_8 / \kappa^\mathrm{e}_\mathrm{K}$ ratio is expected to play only relatively minor role in the uncertainty of the ${\cal M_\mathrm{s}} - \sigma_{N / \langle N \rangle}$ relation. This is because we correlate specifically the \emph{mean-normalized} column density variances with ${\cal M_\mathrm{s}}$. Even though possible use of an erroneous opacity-law between NIR and MIR reflects directly to the column densities, their effect to \emph{mean-normalized} column densities should be a second-order effect. This is  because, to first order, the column densities used in the PDFs (i.e., $A_\mathrm{V} \ge 7$ mag) depend linearly on the opacity-law: for example, increase of  $\kappa_8$ by a factor of 1.5 would lower the maximum column densities approximately by that factor. And the PDF of a mean-normalized variable is invariant to multiplication of the variable by a constant factor. However, the uncertainty of the opacity-law does introduce a second-order effect to the shapes of the PDFs, because the response of column densities to the opacity-law is not \emph{exactly} linear. Below $A_\mathrm{V} \lesssim 10-20 $ mag the large-scale component of the column density field becomes significant compared to the smaller-scale component and the signal of the combined column density mapping technique (by construction, see Section \ref{subsec:combination}) switches from being MIR-dominated to NIR-dominated. We illustrate the response of the combined column densities to the adopted $\kappa_8 / \kappa^\mathrm{e}_\mathrm{K}$ ratio in Fig. \ref{fig:kappa_error} (online only). In this test, we kept the NIR-opacity constant and varied the MIR opacity by 15\%, 30\%, and 60\%. The experiment shows that in these cases the PDF shape of even a mean-normalized column densities below $A_\mathrm{V} \lesssim 10-20$ mag could be altered by the combination procedure. However, we include in the PDFs only column densities above $A_\mathrm{V} \ge 7$ mag, and therefore it seems justified to assume that this kind of bias is clearly a second-order effect. For example, in the test above (which is for Cloud A), the $\sigma_\mathrm{N \ \langle N \rangle}$ changes about 5\% and 20 \% for the cases where MIR opacity is underestimated by 30\% and 60\%, respectively.


Based on the experiments described above, we conclude that the observational data tentatively suggest correlation between the sonic Mach number and  column density dispersion in molecular clouds. \emph{These measurements are, to our knowledge, the first direct measurement of the relationship of these variables and also} the first direct detection of the correlation. However as also described above, we recognize the uncertainty of our measurements, rising mainly from the very small number of clouds in our sample and the uncertainties in defining "the cloud area" in our sample that is not tailored for the measurements of this particular quantity. It would require a larger sample of IRDCs to strengthen this conclusion. In practice, such a study would need to target clouds in locations that are somewhat offset from the Galactic plane and as isolated as possible. 


Finally, we briefly speculate about the interpretation of the slope value we derive from the ${\cal M_\mathrm{s}} - \sigma_{N / \langle N \rangle}$ data.  As noted earlier, \citet{bru10method} examined the 3D-2D variance ratio in a sample of isothermal turbulence simulations and measured the ratios to be between $R=[0.03, 0.15]$. If we simply assume that the ratio is in this range, the mean slope gives the values $b = 0.047 / \sqrt{\{0.03, 0.15\}} = \{0.12, 0.27\}$ with the mean at $b = 0.20$. We report the 3-$\sigma$ confidence range for this mean as the 3-$\sigma$ deviation from the highest and lowest possible values: upper limit is given by  $b = (0.051 + 3 \times 0.016) / \sqrt{0.03} = 0.57$ and the lower limit by $b = (0.042 - 3 \times 0.016) / \sqrt{0.15} = -0.02$. Thus, we report as our measurement $b = 0.20^{+0.37}_{-0.22}$. This result points toward a somewhat lower value of $b$ than derived by \citet{bru10taurus} ($b = 0.49$\footnote{Note that \citet{bru10taurus} reports the value $b = 0.48$, but they use an erroneous Mach number of 17 for Taurus. The correct Mach number for Taurus based on the data they use is ${\cal M_\mathrm{s}} = 10$. \citet{bru10taurus} also uses a correction factor to account for column density variance not included in their measurement. If we do not use the correction factor, in which case their measurement of $\sigma_{N/ \langle N \rangle}$ is very similar to ours, the value $\sigma_\mathrm{N / \langle N \rangle} = 0.84$ follows. Together with their measurement of $R=0.029$ this leads to the value $b = 0.49$.\label{footnote}}) and \citet{pad97apj} ($b \approx 0.5$). Low values of $b$ are generally predicted by (non-magnetized) isothermal simulations in which the turbulence driving is purely solenoidal \citep[b = 1/3, e.g.,][]{fed08b, fed10}, while increasing the fraction of compressive modes increases the value of $b$ toward unity \citep[e.g.,][]{fed10}. 
The presence of magnetic fields also modifies the ${\cal M_\mathrm{s}} - \sigma_\mathrm{\rho / \langle \rho \rangle}$ correlation, and in general, marginally flattens the slope (lowers the $b$) compared to non-magnetized cases \citep[e.g.,][]{pri11}. 
Also, strong density dependence of the magnetic field makes the density fluctuations more weakly dependent on ${\cal M_\mathrm{s}}$ \citep[e.g.,][]{ost01, pri11, mol12}. Even further, when the density dependence of the magnetic field strength approaches $B \propto \rho$, the  ${\cal M_\mathrm{s}} - \sigma_{N / \langle N \rangle}$ is predicted to show \emph{no correlation} at Mach numbers greater than ${\cal M_\mathrm{s}} \gtrsim 7$ \citep{mol12}. In this context, the fact that we detect a ${\cal M_\mathrm{s}} - \sigma_\mathrm{N / \langle N \rangle}$ correlation up to high Mach numbers supports the picture in which the magnetic field depends relatively weakly on density. This has also been suggested through Zeeman effect observations \citep[$B \propto \rho^{0.65}$ at $n > 300$ cm$^{-3}$,][]{cru10}. 

\subsection{Comparison with earlier work} 


How do our results regarding the ${\cal M_\mathrm{s}} - \sigma_\mathrm{\rho / \langle \rho \rangle}$ relation compare with the previous works on the topic? The ${\cal M_\mathrm{s}} - \sigma_\mathrm{\rho / \langle \rho \rangle}$ relation (or ${\cal M_\mathrm{s}} - \sigma_{N / \langle N \rangle}$ relation) has not been widely examined by observations. The most comprehensive work to our knowledge is that of \citet{bru10taurus} who exploit CO line emission data and near-infrared dust extinction data to estimate ${\cal M_\mathrm{s}}$ and $\sigma_{N / \langle N \rangle}$, respectively, and use the method of \citet{bru10method} to infer $\sigma_\mathrm{\rho / \langle \rho \rangle}$ and the proportionality constant $b$. Their observations indicate the value of $b=0.49$ (see Footnote \ref{footnote}). A similar technique was applied earlier to the dust extinction data of IC5146 by \citet{pad97apj} who estimated the value $b \approx 0.5$. \emph{Both of these works, however, only measure one ${\cal M_\mathrm{s}} - \sigma_{N / \langle N \rangle}$ ratio pair, and not explicitly the correlation between the two variables.}

\begin{figure*}
   \centering
\includegraphics[width=\columnwidth]{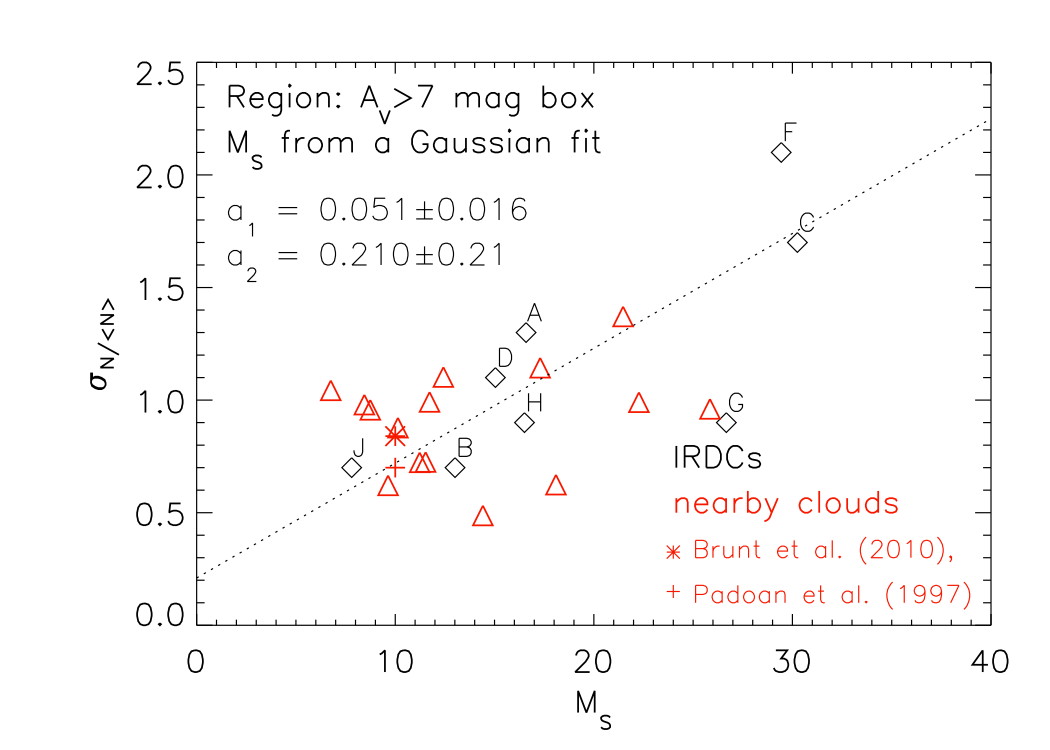}\includegraphics[width=\columnwidth]{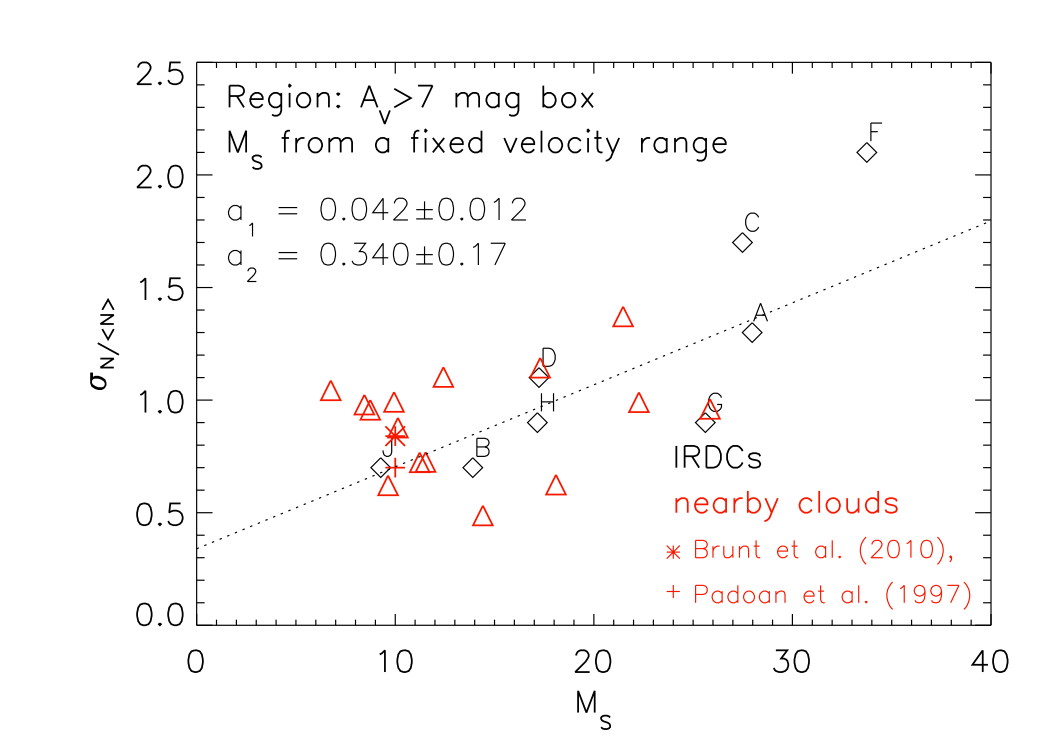}
      \caption{Standard deviation of the mean-normalized column densities as a function of the sonic Mach number in eight IRDCs (black diamonds) and 14 nearby molecular clouds (red triangles). To compute the $\sigma_{N / \langle N \rangle}$ values of the IRDCs, we fitted log-normal functions to the PDFs of the clouds and used the mean column density resulting from the fit to normalize the column density values. The PDF was constructed from the box-like area around the cloud as explained in Section \ref{subsec:pdfs}. The $\sigma_{N / \langle N \rangle}$ values of the nearby clouds are calculated over a dynamic range that should represent well the total column density dispersion (see text). The value derived by \citet{bru10taurus} and \citet{pad97apj} is shown with a red asterisk. The dotted lines show linear fits to all data points. \emph{Left: }${\cal M_\mathrm{s}}$ is calculated from Gaussian fits to the mean spectra. \emph{Right: }${\cal M}_\mathrm{s}$ is calculated from a fixed velocity range (see Section \ref{subsec:co}).
                    }
         \label{fig:var_vs_var2}
   \end{figure*}

\citet{goo09} examined the PDFs and variances of the column density field in six sub-regions of the Perseus molecular cloud. The data (and the method) used by their work were very similar to what we employ in this paper. They tested if ${\cal M_\mathrm{s}}$ increases with increasing 2D variance (and mean) and they conclude that it does not. Thus, the result of \citet{goo09} suggests no correlation between $\sigma_{\rho / \langle \rho \rangle}$ and ${\cal M_\mathrm{s}}$. 
%
%
In the scope of the data we show in Fig. \ref{fig:var_vs_var}, it seems clear that the contradiction between our and \citet{goo09} results originate from the fact that the velocity dispersions in the sub-regions within Perseus span only a narrow range (Goodman et al. do not quote the linewidths of each sub-region, but the integrated linewidth over the entire Perseus is $\approx 1$ km s$^{-1}$, inevitably restricting the values to a rather narrow range). This leaves little leverage for the correlation to be detected, given the large scatter in the data shown in Fig. \ref{fig:var_vs_var2}. 


\section{Conclusions} 
\label{sec:conclusions}

We present in this paper a new technique for probing the column density structure of IRDCs. The technique is based on combining column density maps derived from NIR observations of background stars shining through the IRDCs (\emph{UKIDSS/Galactic Plane Survey data) and from MIR (8 $\mu$m) shadowing features against the Galactic background (\emph{Spitzer/GLIMPSE} survey)}. We demonstrate the use of the method by deriving column density data for ten IRDC complexes. We then use the resulting maps to examine the relation between the sonic Mach number and column density variance in the clouds. The main conclusions of our work are as follows.
	
\begin{enumerate}

   \item The technique we present provides temperature-independent column density data over a unique combination of spatial resolution ($\sim$2$\arcsec$) and dynamic range ($A_\mathrm{V} \approx 1-100$ mag). Unlike the mid-infrared technique alone, the new method is well-calibrated at column densities (1 mag $< A_\mathrm{V} \lesssim 10$ mag). The new information on the low-column density surroundings of the IRDCs can significantly improve our understanding of the impact of the envelope material to the physics of the IRDCs \citep[e.g.][]{her12}. The technique is applicable to clouds up to about $D \lesssim 8$ kpc, thereby potentially reaching a great number of IRDCs. 
   
   \item The column density PDFs of the IRDCs in the range $A_\mathrm{V} = 7-100$ mag are, on average, consistent with a wide log-normal shape that has a width of $\sigma_\mathrm{\ln{N}} \approx 0.9$. This shape is quite similar, although slightly shallower, to the PDFs of nearby active star-forming clouds reported by K09 in the column density range in which the data overlap, i.e., $A_\mathrm{V} = 7-25$ mag. 
   
   \item The IRDCs contain relatively high amount of high-column density material, comparable to the fraction observed in nearby active star-forming clouds such as Orion A, and possibly even exceeding it. The average cumulative distribution function of IRDCs between $7 \mathrm{\ mag} < A_\mathrm{V} < 100$ mag behaves exponentially, $CDF \propto e^{-0.07 \times N}$. In the framework of the recent work on the star-forming rates of molecular clouds \citep{lad12}, this suggests that the star-forming rates (per unit cloud mass) of IRDCs are at least comparable to active nearby clouds.
   
   \item We present the first direct observational measurement of the relation between the turbulent energy (sonic Mach number, ${\cal M_\mathrm{s}}$), and density fluctuations (column density dispersion, $\sigma_{N / \langle N \rangle}$), in molecular clouds. We report a tentative detection of correlation with about 3-$\sigma$ confidence. Linear model fits to the $({\cal M_\mathrm{s}}, \sigma_{N / \langle N \rangle})$ data yields the slopes of $\sigma_{N / \langle N \rangle} = (\{0.051, 0.042\} \pm \{0.016, 0.012 \} )\times {\cal M_\mathrm{s}}$ for the two methods we use for measuring ${\cal M_\mathrm{s}}$. 

   \item Our results suggest the correlation coefficient $b \approx 0.20^{+0.37}_{-0.22}$ between the sonic Mach number and volume density variance, $\sigma_\mathrm{\rho / \langle \rho \rangle}$, in molecular clouds. In this expression, the quoted uncertainties indicate the 3-$\sigma$ confidence interval assuming that the 2D-3D variance ratio $R$ is between $R = [0.03, 0.15]$ \citep[as measured by][]{bru10method}. The value of $b$ we derive is in agreement with earlier derivations by \citet{pad97apj} and \citet{bru10taurus}, that are, however, based on measurements of only single $({\cal M_\mathrm{s}}, \sigma_{N / \langle N \rangle})$ ratio pairs. 

   \item When combined with recent numerical and analytical predictions \citep{pri11, mol12}, the detection of correlation between ${\cal M_\mathrm{s}}$ and $\sigma_{N / \langle N \rangle}$ is suggestive of relatively weak density dependence of the magnetic field strength in the clouds. This is in agreement with the field strength measurements of \citet{cru10}.

\end{enumerate}

While the conclusions given above remain tentative, the technique we present provides a good basis for measuring the relationship between ${\cal M_\mathrm{s}} - \sigma_{N / \langle N \rangle}$ in a statistically significant sample of molecular clouds in the future.


\begin{acknowledgements}
The authors are grateful to Henrik Beuther, Blakesley Burkhart, Christoph Federrath, Simon Glover, Thomas Henning, Ralf Klessen, and Lukas Konstandin for fruitful discussions and comments regarding the manuscript. We thank the referee whose comments improved the manuscript significantly.
The work of JK was supported by the Deutsche Forschungsgemeinschaft priority program 1573 ("Physics of the Interstellar Medium"). 	
JK gratefully acknowledges support from The Finnish Academy of Science and Letters, V\"ais\"al\"a Foundation for this work. 
JCT acknowledges support from NSF CAREER grant AST-0645412; NASA Astrophysics Theory and Fundamental Physics grant ATP09-0094 and NASA Astrophysics Data Analysis Program ADAP10-0110. This work was assisted by the ASTROWIN workshop program supported by the U. Florida Dept. of Astronomy.
This publication makes use of molecular line data from the Boston University-FCRAO Galactic Ring Survey (GRS). The GRS is a joint project of Boston University and Five College Radio Astronomy Observatory, funded by the National Science Foundation under grants AST-9800334, AST-0098562, \& AST-0100793.  
\end{acknowledgements}



\newpage

\appendix

\section{CO spectra towards the IRDCs}
\label{app:co}


   \begin{figure*}
   \centering
\includegraphics[width=0.33\textwidth]{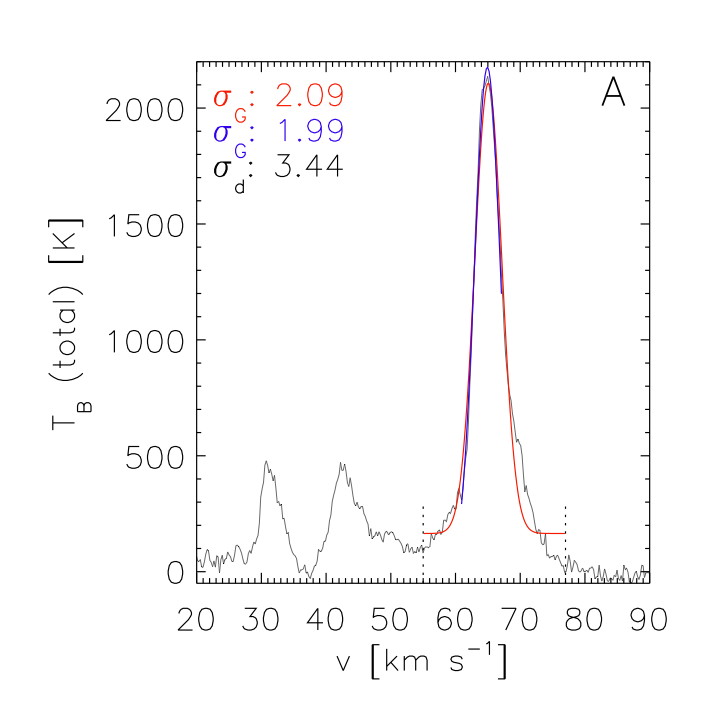}\includegraphics[width=0.33\textwidth]{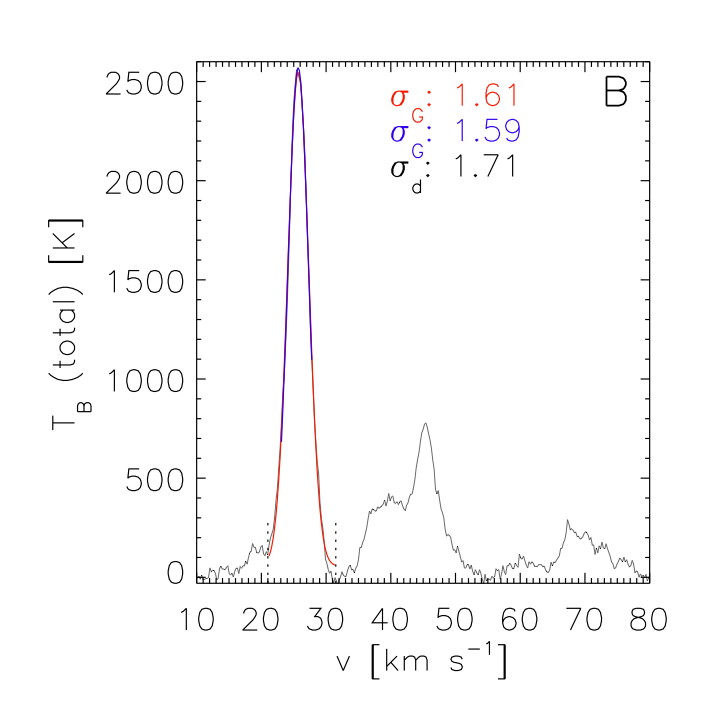}\includegraphics[width=0.33\textwidth]{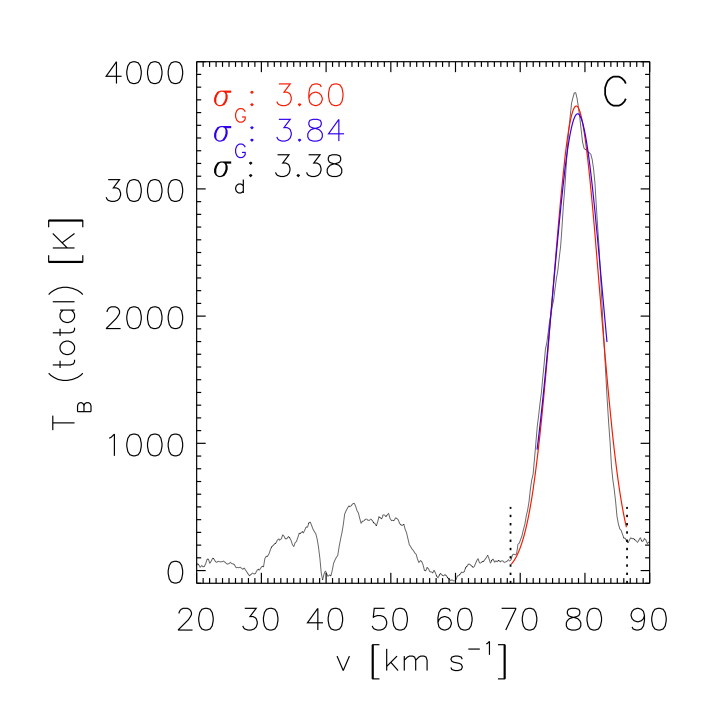}
\includegraphics[width=0.33\textwidth]{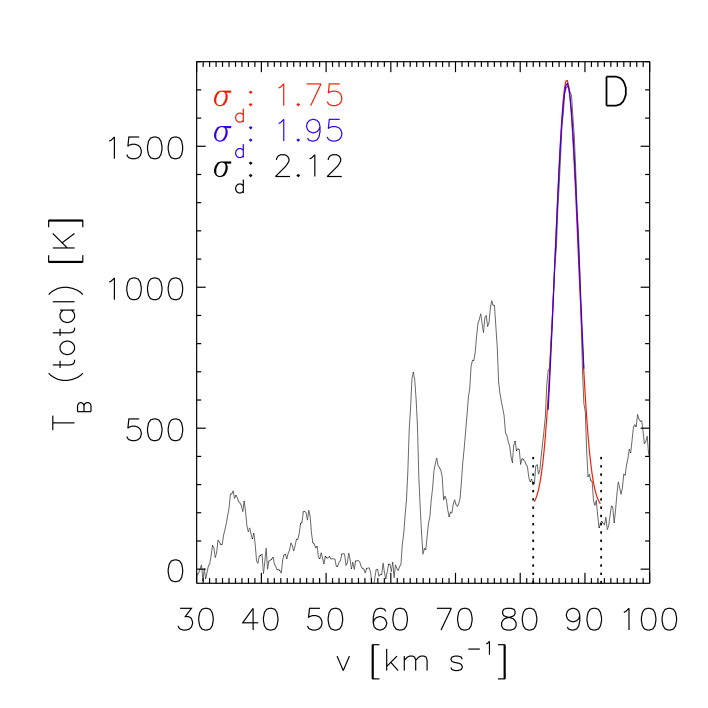}\includegraphics[width=0.33\textwidth]{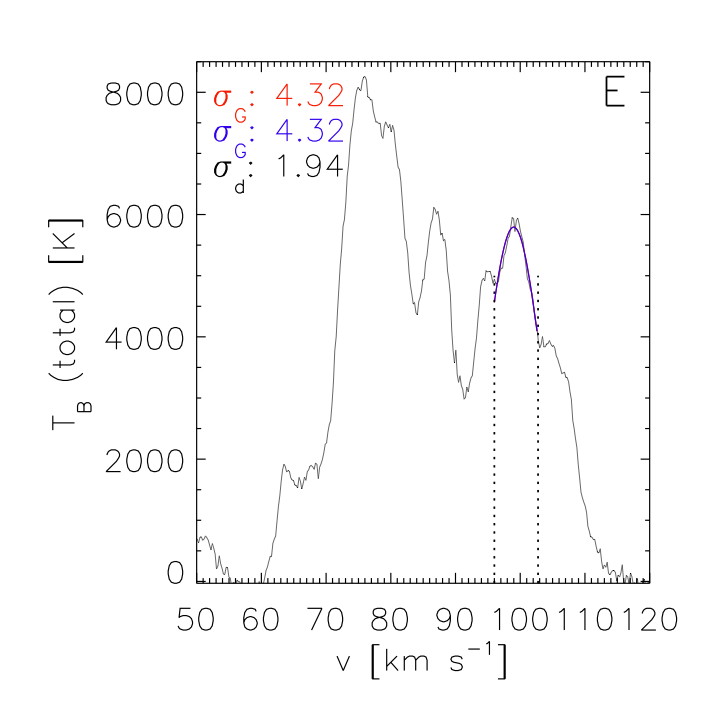}\includegraphics[width=0.33\textwidth]{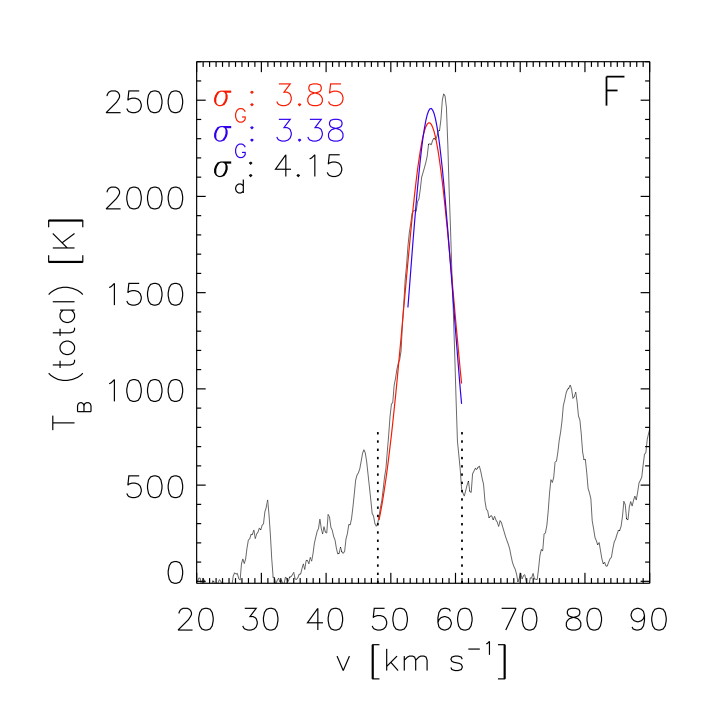}
\includegraphics[width=0.33\textwidth]{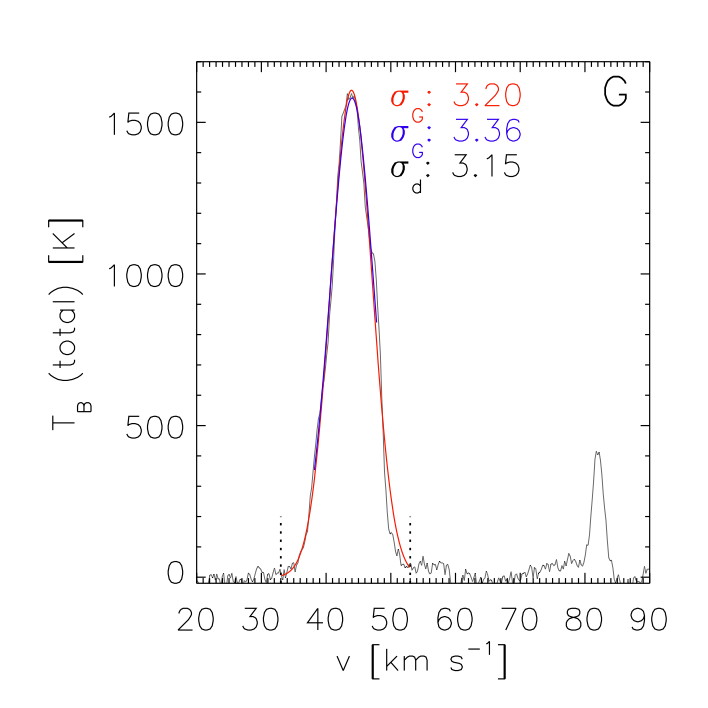}\includegraphics[width=0.33\textwidth]{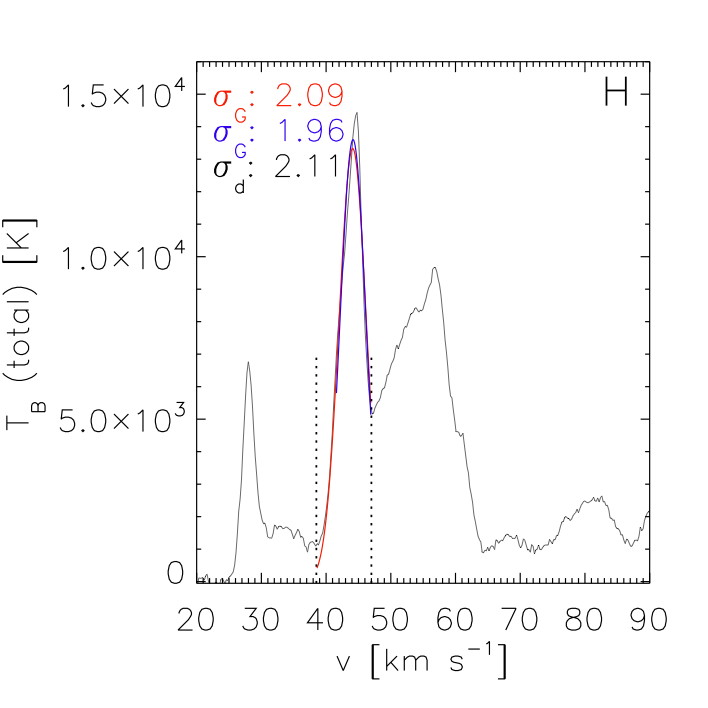}\includegraphics[width=0.33\textwidth]{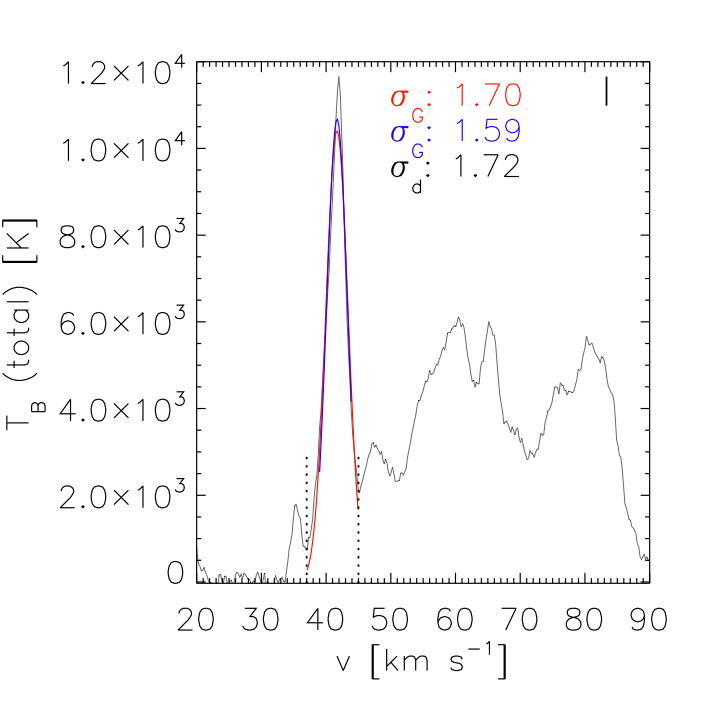}
\includegraphics[width=0.33\textwidth]{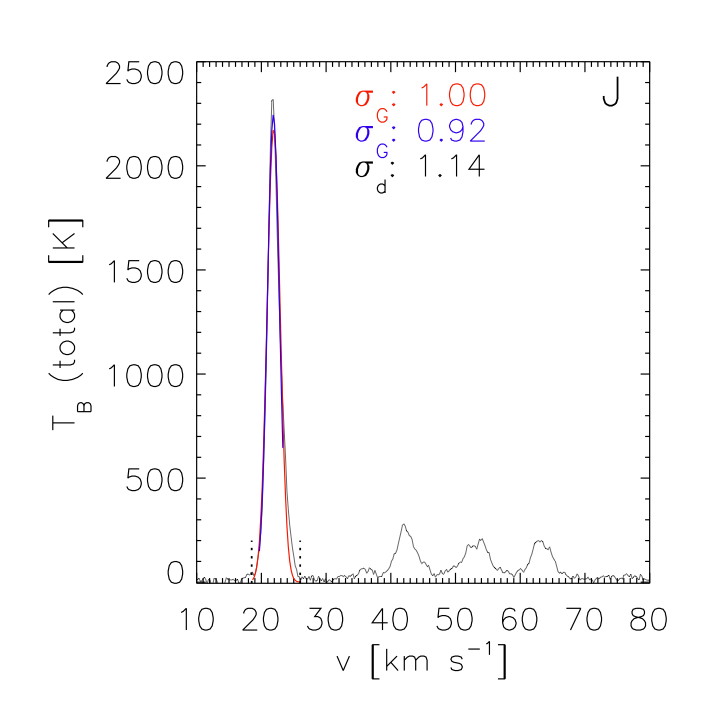}
      \caption{The mean $^13$CO spectra of the IRDCs of our sample. The spectra have been averaged over the $A_\mathrm{V} > 7$ mag box (see Section \ref{subsec:area} for the detailed definition). The dotted vertical lines indicate the velocity interval chosen to represent the cloud. The red line shows a fit of a Gaussian to this velocity interval. The blue line shows another Gaussian fit, performed over the interval $v_\mathrm{peak} - 1.5\sigma, v_\mathrm{peak} = 1.5\sigma$ where $\sigma$ is the dispersion from the first Gaussian fit. The dispersions are shown in the panels. The third dispersion value, $\sigma_\mathrm{d}$ gives the standard deviation of the data within the chosen velocity interval.
              }
         \label{fig:co_box}
   \end{figure*}


   \begin{figure*}
   \centering
\includegraphics[width=0.33\textwidth]{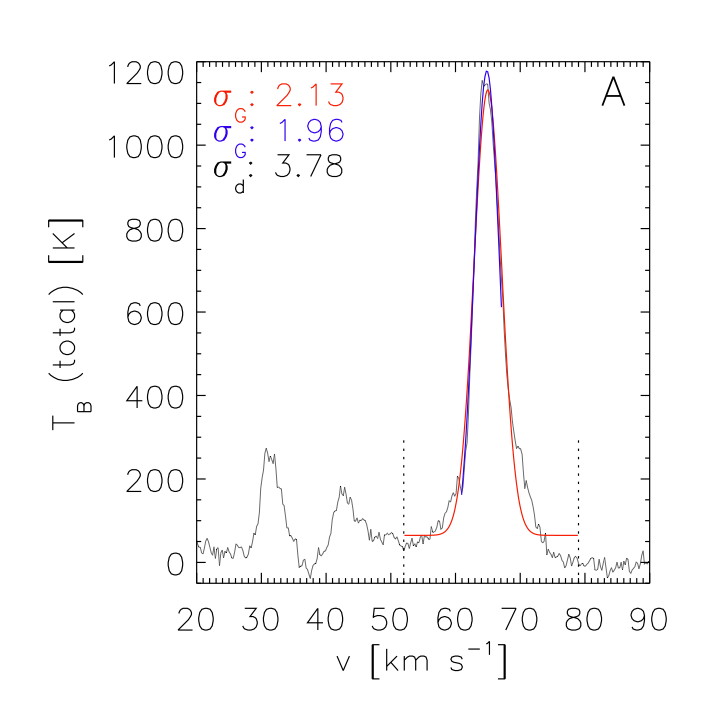}\includegraphics[width=0.33\textwidth]{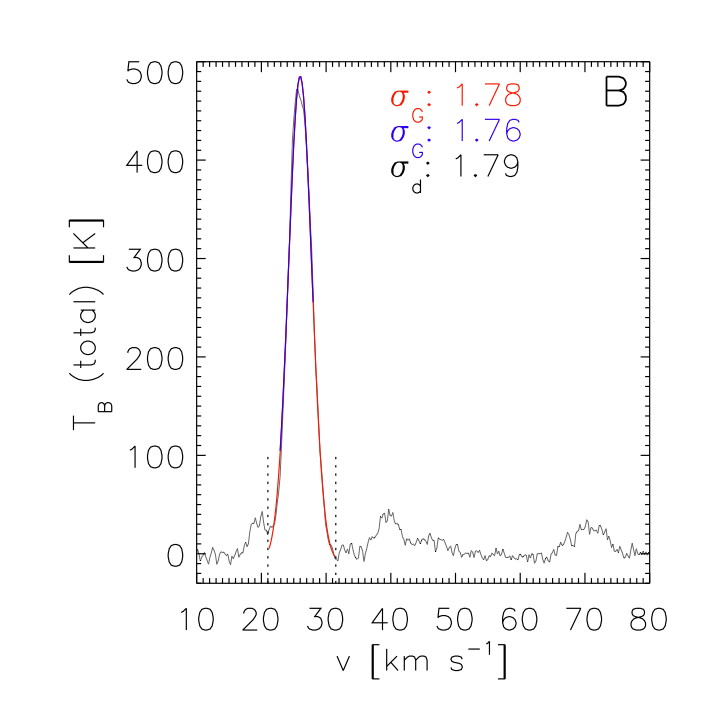}\includegraphics[width=0.33\textwidth]{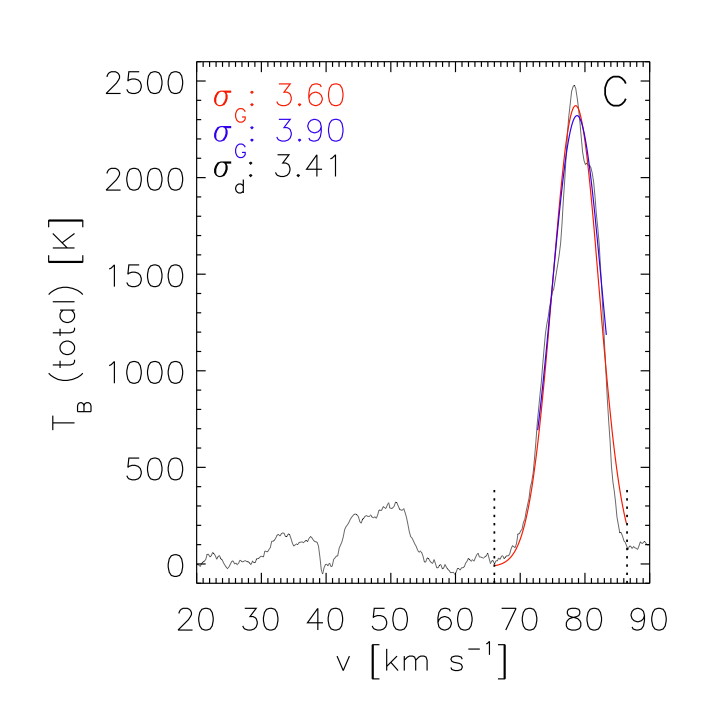}
\includegraphics[width=0.33\textwidth]{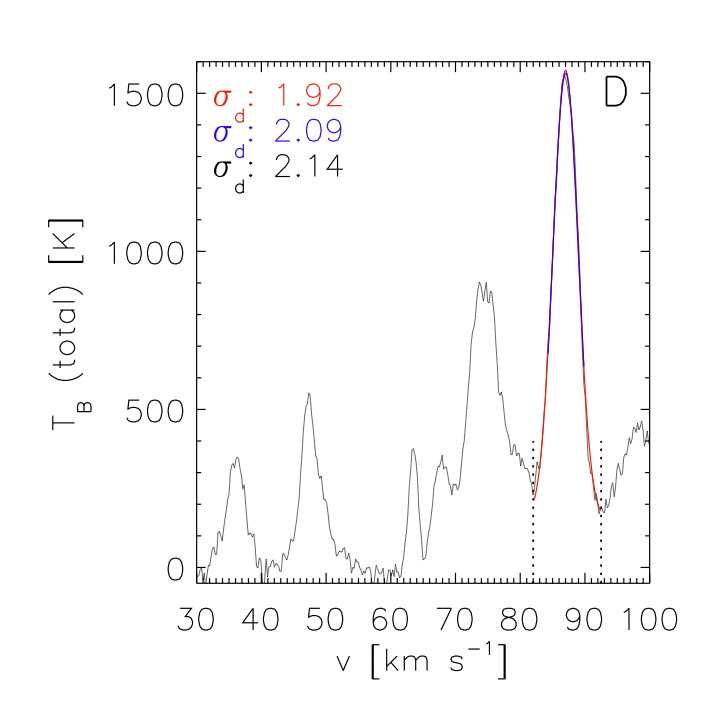}\includegraphics[width=0.33\textwidth]{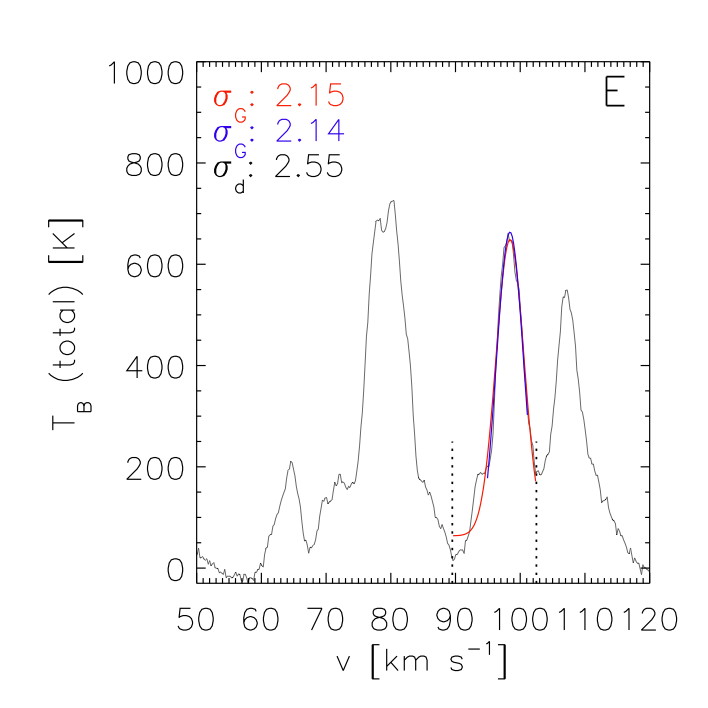}\includegraphics[width=0.33\textwidth]{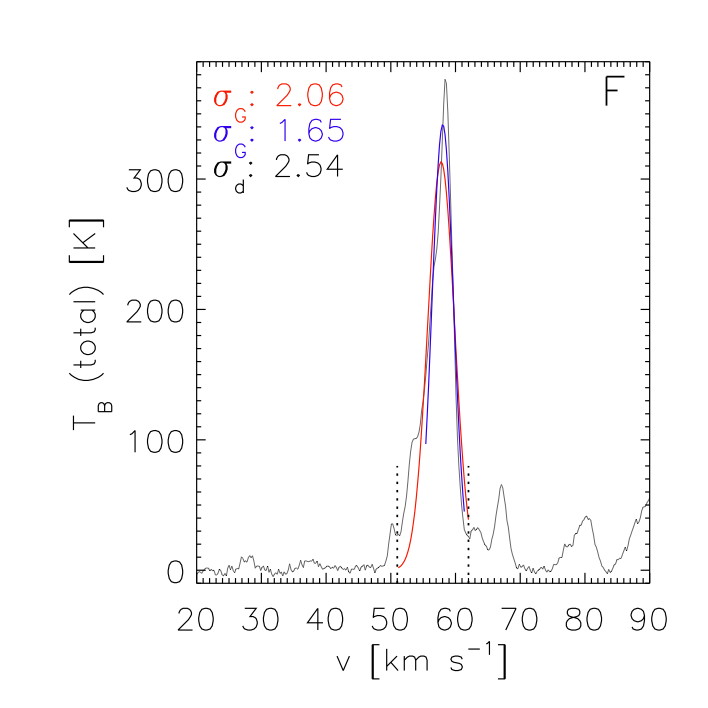}
\includegraphics[width=0.33\textwidth]{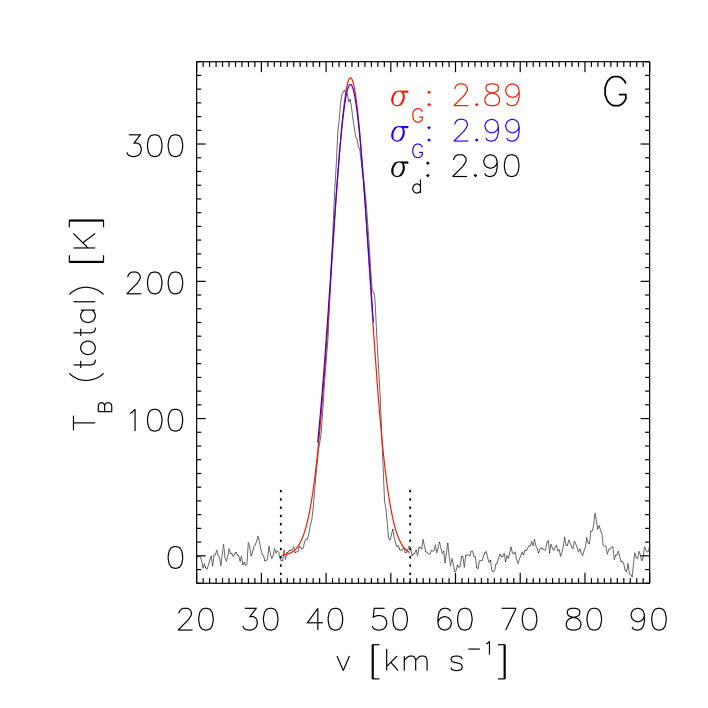}\includegraphics[width=0.33\textwidth]{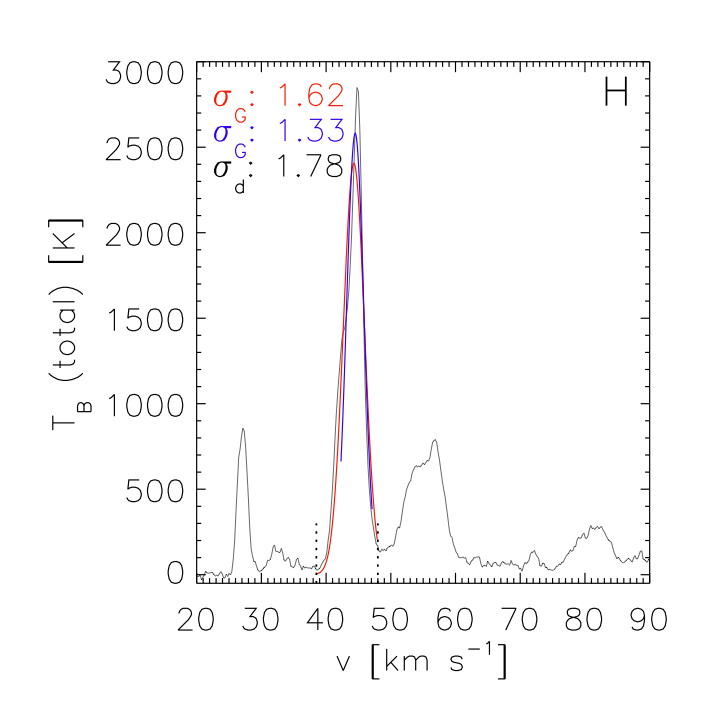}\includegraphics[width=0.33\textwidth]{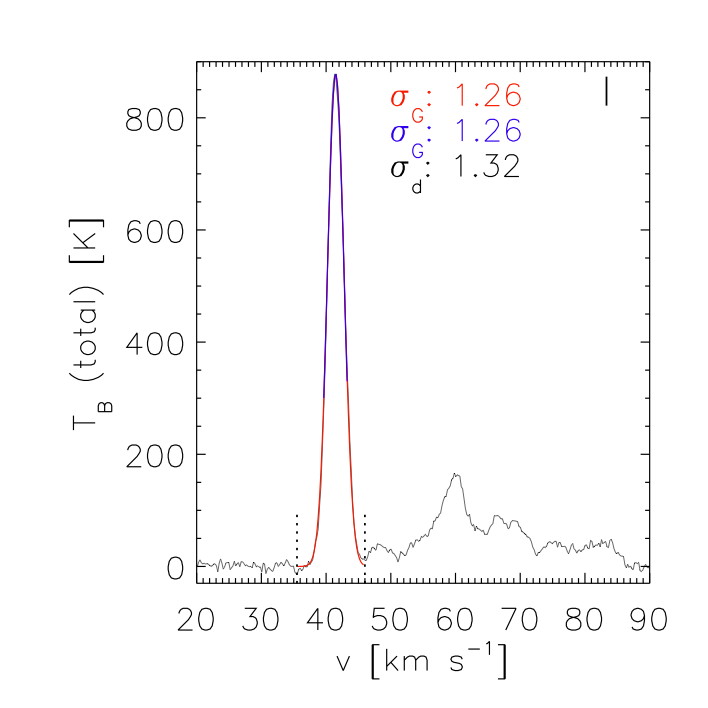}
\includegraphics[width=0.33\textwidth]{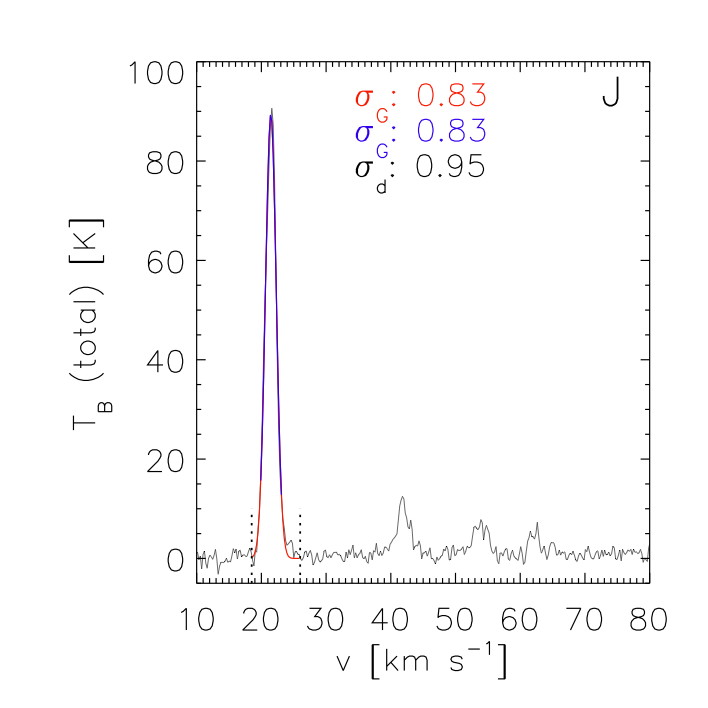}
      \caption{The same as Fig. \ref{fig:co_box}, but the spectra have been averaged over the \citet{sim06} ellipsoids (see Section \ref{subsec:area} for the detailed definition).
              }
         \label{fig:co_ell}
   \end{figure*}

\section{High-dynamic-range column density maps of the IRDC sample}
\label{app:data}

   \begin{figure*}
   \centering
\includegraphics[bb = 00 0 800 400, clip=true, width=1.3\textwidth]{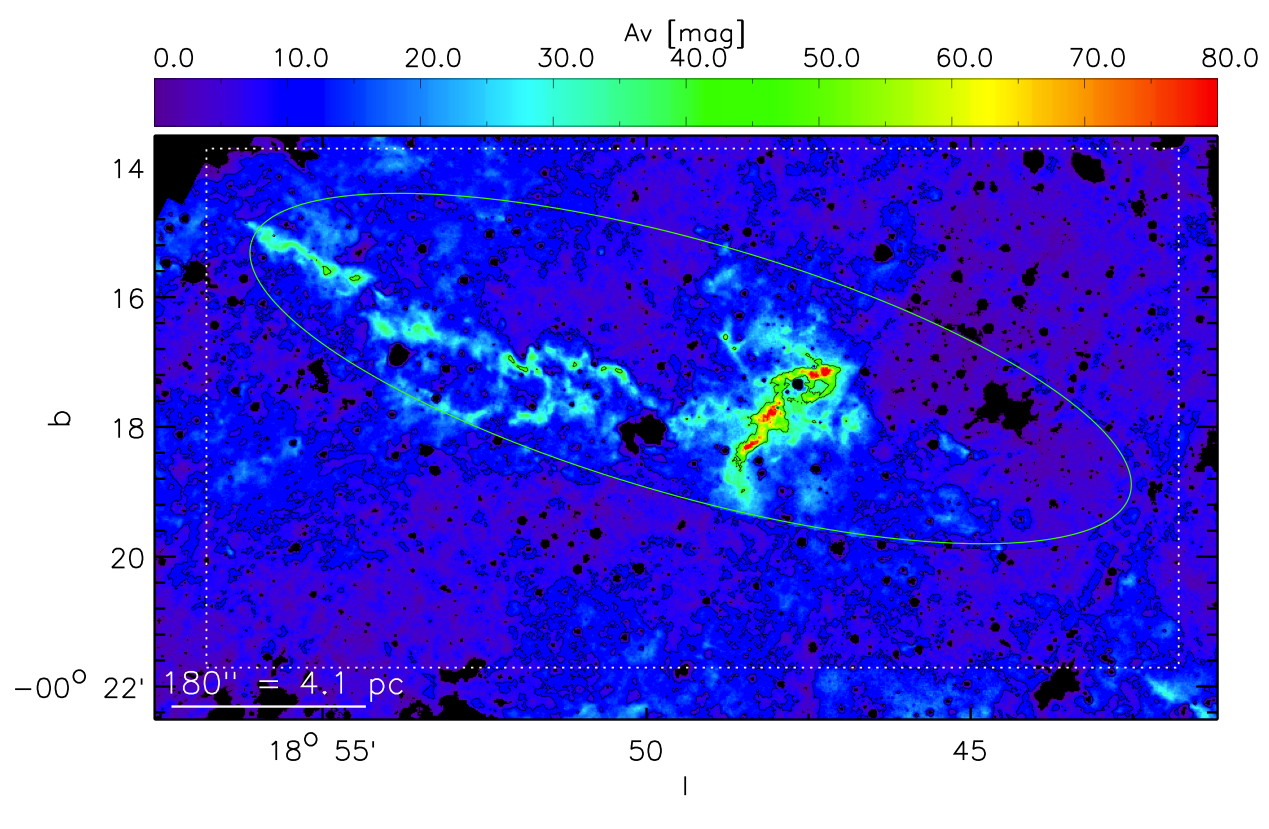}
      \caption{High-dynamic-range column density map of the cloud A, derived using a combination of MIR and NIR data. The green circle shows the largest ellipse from the \citet{sim06} catalog in the region. The white box outlines the region which was used alongside with the $A_\mathrm{V} = 7$ mag contour to define the region that is included in the analyses presented in Section \ref{sec:results}. The contours are drawn at $A_\mathrm{V} = [7, 40]$ mag. The scale bar shows the physical scale assuming the distance as given in Table \ref{tab:clouds}.  
              }
         \label{fig:maps-A}
   \end{figure*}

   \begin{figure*}
   \centering
\includegraphics[bb = 120 0 700 430, clip=true, width=1.3\textwidth]{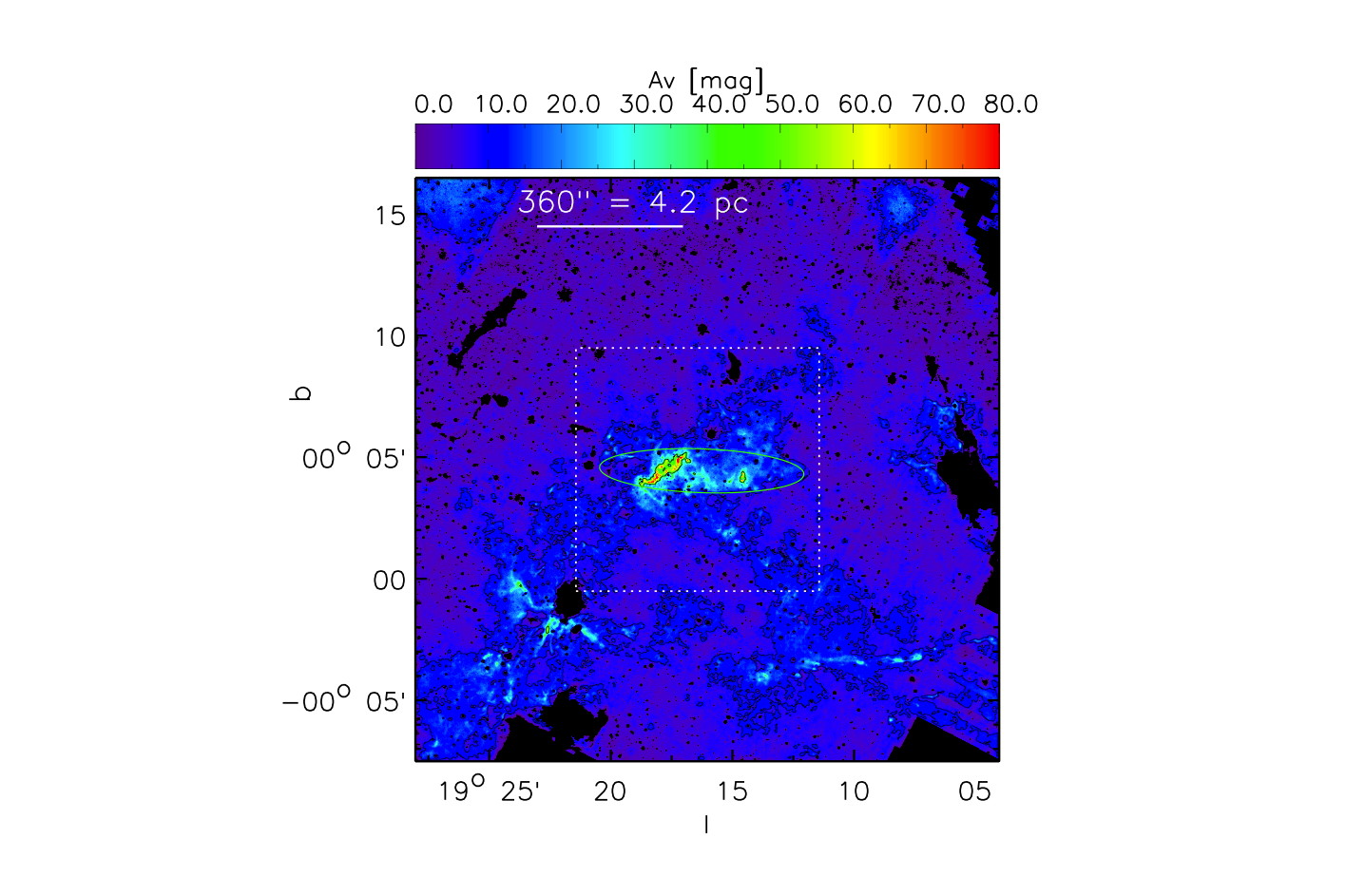}
      \caption{Same as Fig. \ref{fig:maps-A}, but for the cloud B.
              }
         \label{fig:maps-B}
   \end{figure*}

   \begin{figure*}
   \centering
\includegraphics[bb = 120 0 700 430, clip=true, width=1.3, width=1.3\textwidth]{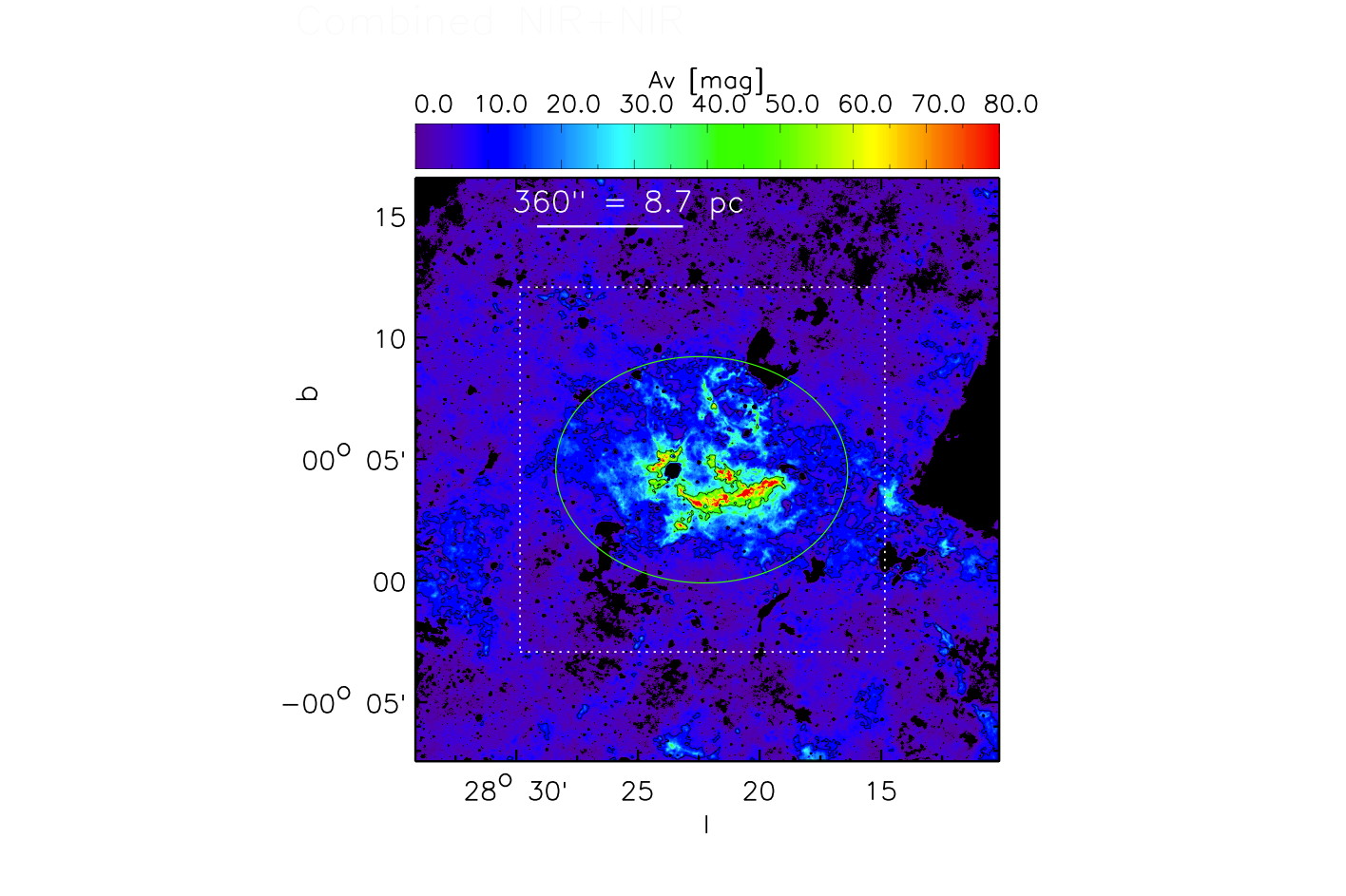}
      \caption{Same as Fig. \ref{fig:maps-A}, but for the cloud C.
              }
         \label{fig:maps-C}
   \end{figure*}

   \begin{figure*}
   \centering
\includegraphics[bb = 120 0 700 430, clip=true, width=1.3, width=1.3\textwidth]{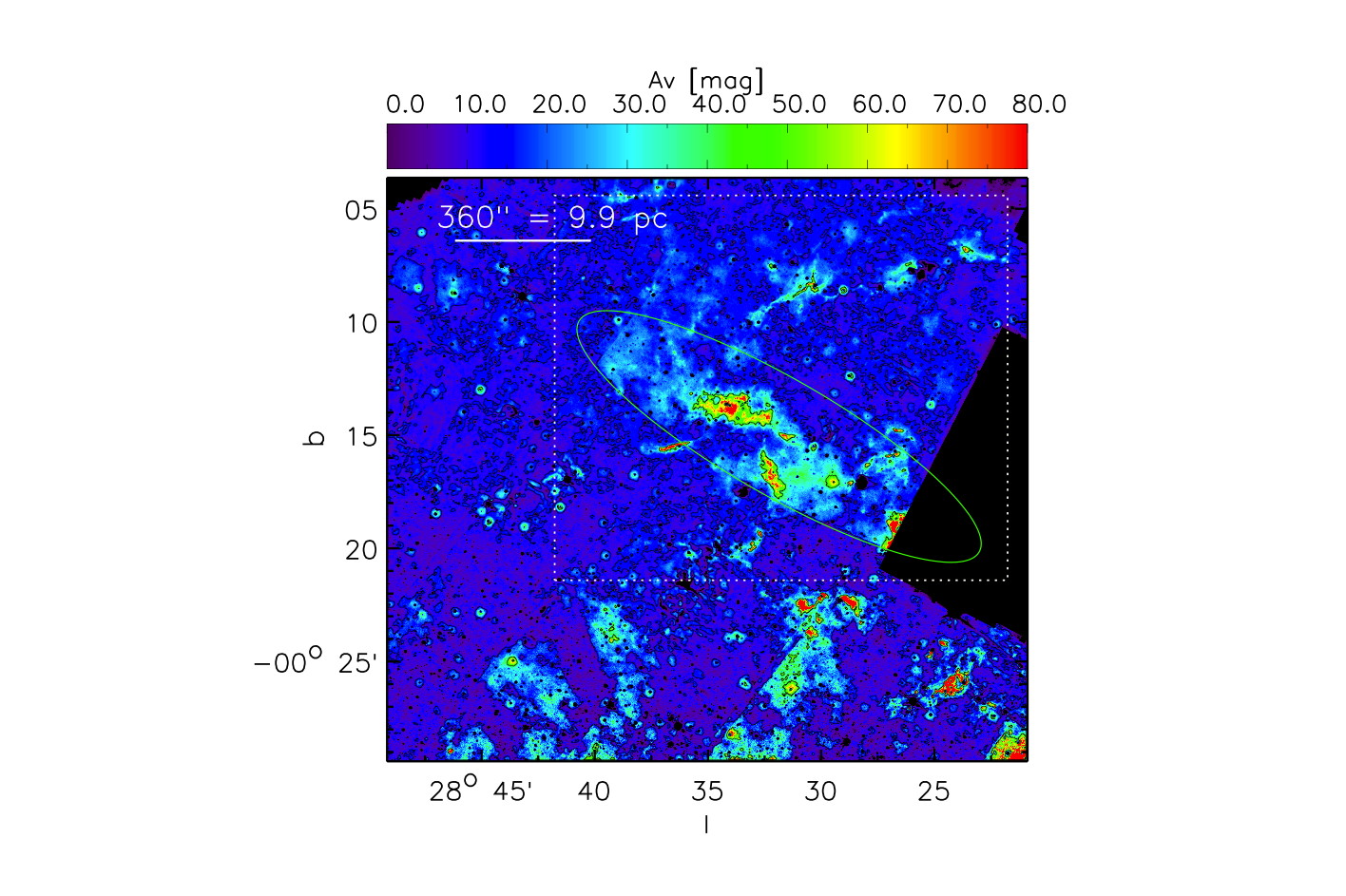}
      \caption{Same as Fig. \ref{fig:maps-A}, but for the cloud D. The rectangular empty area results from missing NIR data. 
              }
         \label{fig:maps-D}
   \end{figure*}

   \begin{figure*}
   \centering
\includegraphics[bb = 120 0 700 430, clip=true, width=1.3, width=1.3\textwidth]{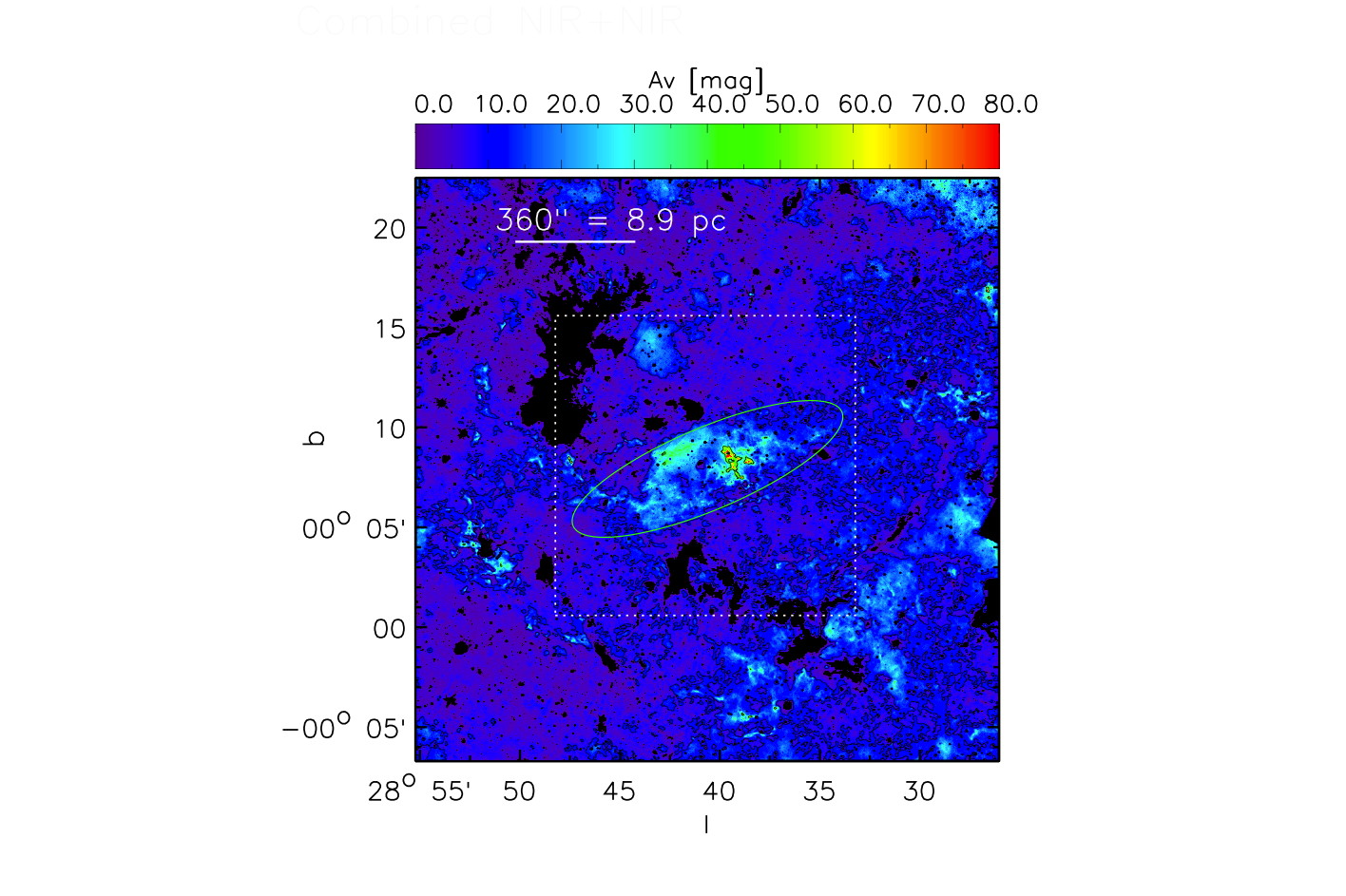}
      \caption{Same as Fig. \ref{fig:maps-A}, but for the cloud E.
              }
         \label{fig:maps-E}
   \end{figure*}

   \begin{figure*}
   \centering
\includegraphics[bb = 120 0 700 430, clip=true, width=1.3, width=1.3\textwidth]{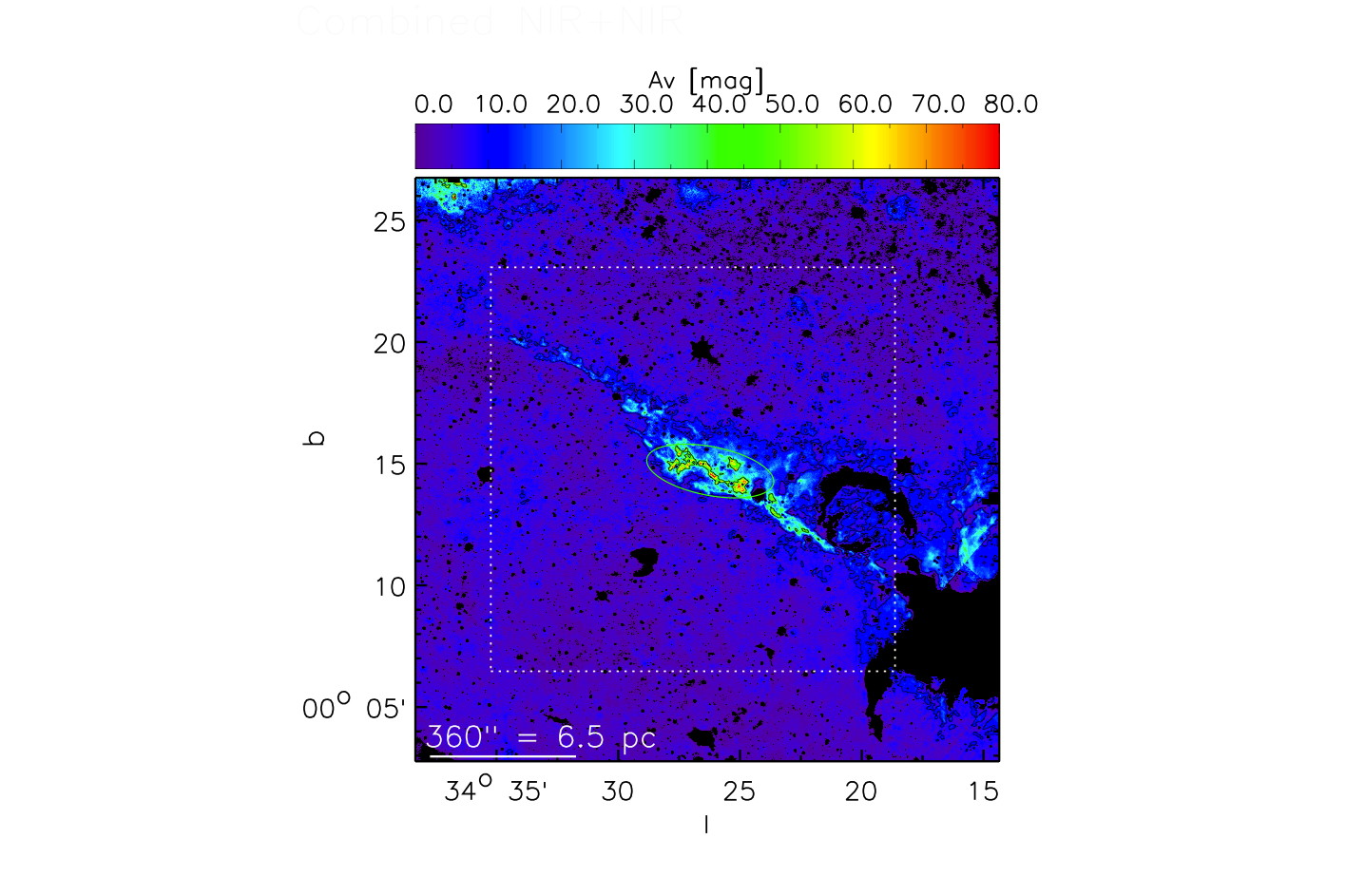}
      \caption{Same as Fig. \ref{fig:maps-A}, but for the cloud F. The areas of missing data (marked with zeros) result from strong MIR nebulosity that hinders the MIR mapping technique.
              }
         \label{fig:maps-F}
   \end{figure*}

   \begin{figure*}
   \centering
\includegraphics[bb = 120 0 700 430, clip=true, width=1.3, width=1.3\textwidth]{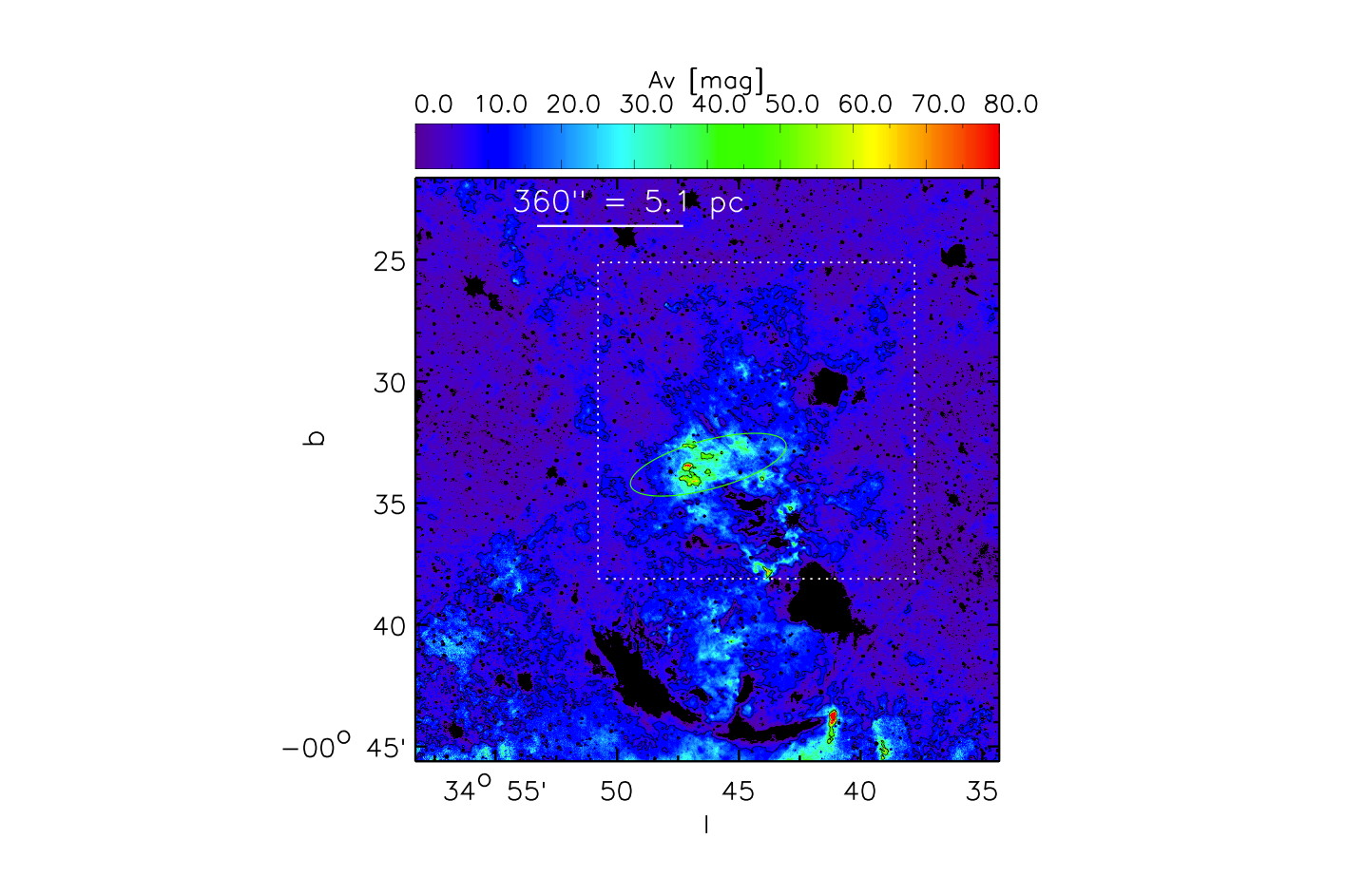}
      \caption{Same as Fig. \ref{fig:maps-A}, but for the cloud G.
              }
         \label{fig:maps-G}
   \end{figure*}

   \begin{figure*}
   \centering
\includegraphics[bb = 100 0 700 430, clip=true, width=1.3, width=1.3\textwidth]{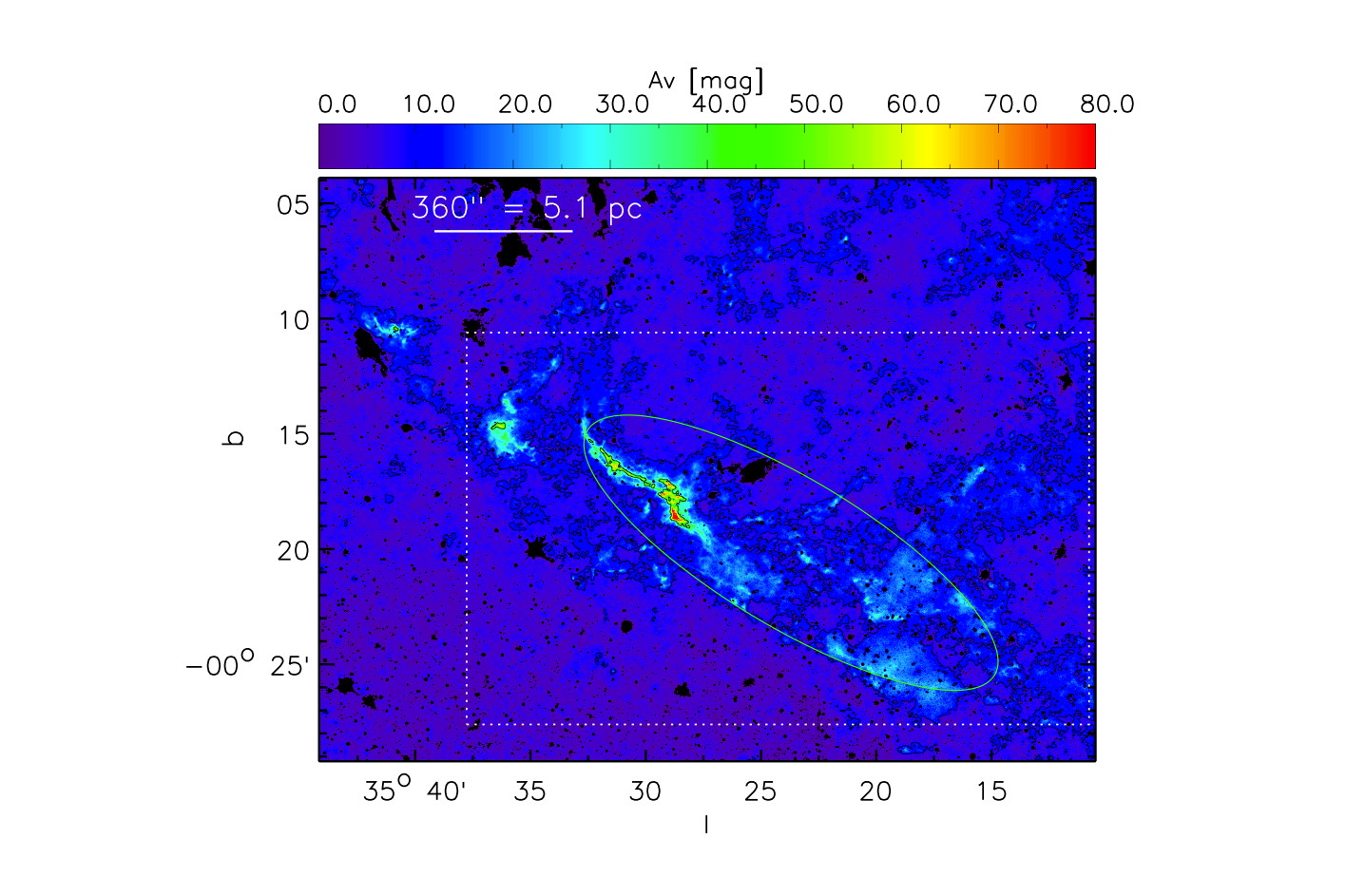}
      \caption{Same as Fig. \ref{fig:maps-A}, but for the cloud H.
              }
         \label{fig:maps-H}
   \end{figure*}

   \begin{figure*}
   \centering
\includegraphics[bb = 120 0 700 430, clip=true, width=1.3, width=1.3\textwidth]{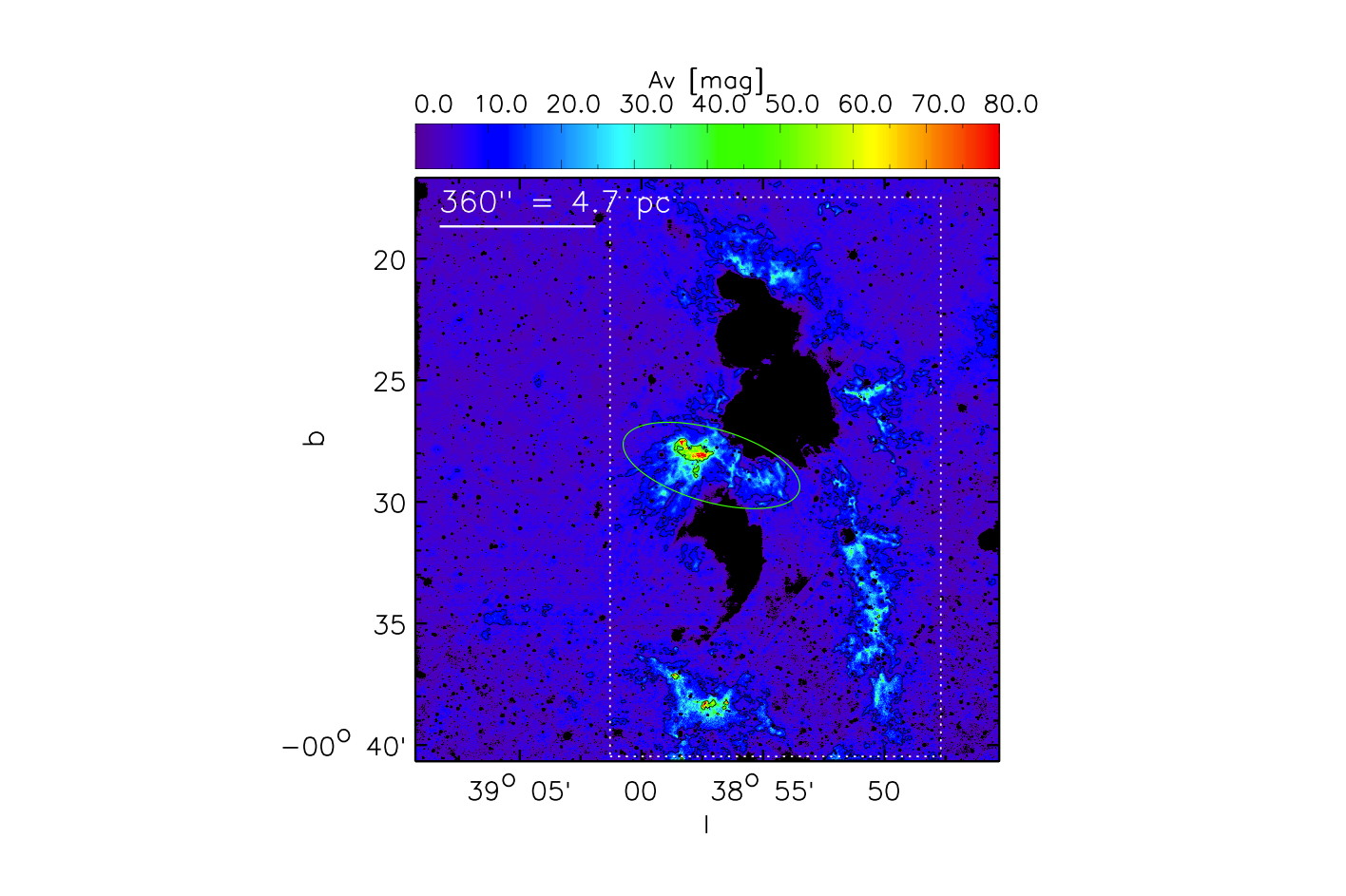}
      \caption{Same as Fig. \ref{fig:maps-A}, but for the cloud I. The areas of missing data (marked with zeros) result from strong MIR nebulosity that hinders the MIR mapping technique.
              }
         \label{fig:maps-I}
   \end{figure*}

   \begin{figure*}
   \centering
\includegraphics[bb = 120 0 700 430, clip=true, width=1.3, width=1.3\textwidth]{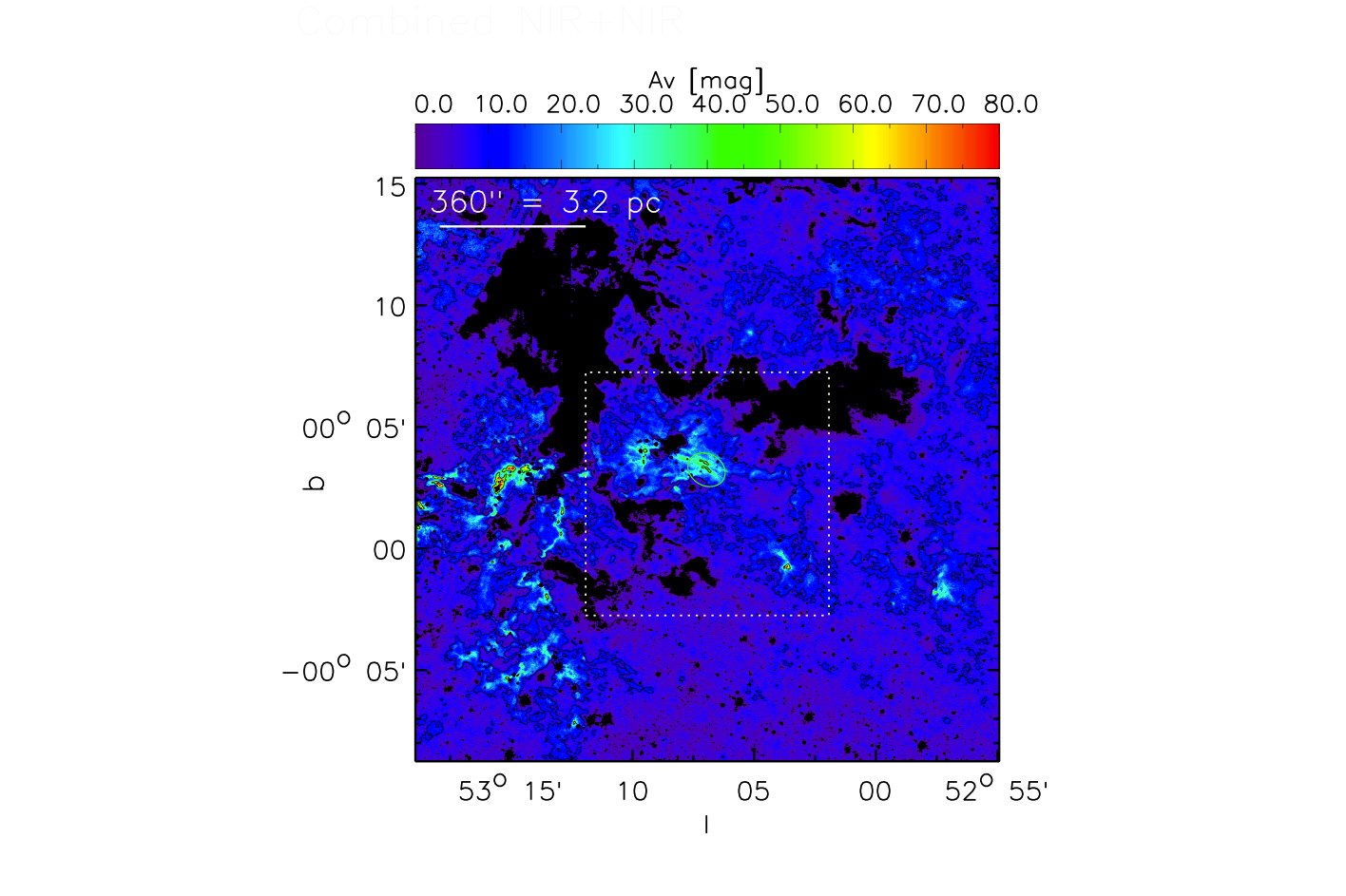}
      \caption{Same as Fig. \ref{fig:maps-A}, but for the cloud J. The areas of missing data (marked with zeros) result from strong MIR nebulosity that hinders the MIR mapping technique.
              }
         \label{fig:maps-J}
   \end{figure*}

\section{Response of the technique to changes of the opacity-law}
\label{app:kappa_error}

   \begin{figure*}
   \centering
\includegraphics[width=\textwidth]{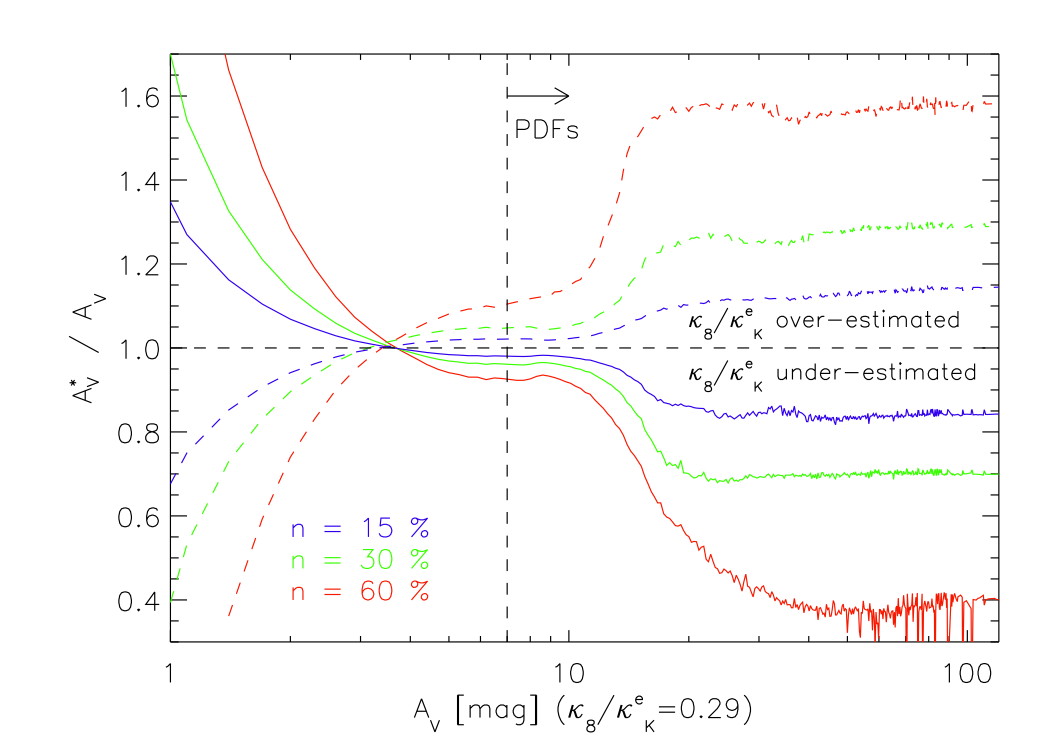}
      \caption{Response of the column densities derived using the combined NIR+MIR technique to the adopted NIR-to-MIR opacity-law. The curves show the ratio of column densities derived with alternative opacity-laws ($\kappa_\mathrm{8} / \kappa^\mathrm{e}_\mathrm{K} = n \times 0.29$) to the one adopted in this paper ($\kappa_\mathrm{8} / \kappa^\mathrm{e}_\mathrm{K} = 0.29$). The curves were calculated by keeping the NIR opacity fixed and by changing the relative MIR opacity. The solid blue, green, and red curves show the case in which the true opacity-law is 15\%, 30\%, and 60\% higher than the adopted value ($\kappa_\mathrm{8} / \kappa^\mathrm{e}_\mathrm{K} =0.29$). The dashed curves show the corresponding curves for the cases in which the ratio is lower. At $A_\mathrm{V} \lesssim 10$ mag the technique relies dominantly on the NIR data, and therefore the changes in MIR opacity do not greatly affect the column density. The large-scale background component that is filtered out by the MIR data (and assumed to be recovered by the NIR data) has its maximum at $A_\mathrm{V} = 10-20$ mag. Thus, at that range there is a transition from NIR-dominated to MIR-dominated regime. Finally at $A_\mathrm{V} \gtrsim 20$ mag the large-scale column density component starts to be small compared to the total column density, and thus the ratio approaches the value by which the original opacity law was modified (i.e., factors 1.15, 1.3, and 1.6). At extinctions lower than the 10 mag threshold value, there is a small bias in the extinctions. It is caused by the fact that the background correction value (see Section \ref{subsec:combination}) for some pixels can result from interpolation from neighboring pixels instead of from the difference of the NIR and MIR extinctions. The pixels for which this interpolation is performed are not necessarily the same in the cases where $n = 1$ and $n \ne 1$.
              }
         \label{fig:kappa_error}
   \end{figure*}

\end{document}